\documentclass[3p,times]{elarticle}

\usepackage{url}
\usepackage{hyperref}
\usepackage{amsmath}

\usepackage{ecr}

\volume{00}

\firstpage{1}

\runauth{L.I. Gurvits et al.}

\jid{procs}

\usepackage{amssymb}

\usepackage[figuresright]{rotating}

\usepackage{graphicx}
\usepackage{xcolor} 
\usepackage[normalem]{ulem} 

\usepackage{siunitx} 

\begin{document}

\begin{frontmatter}



\dochead{}

\title{The science case and challenges of space-borne sub-millimeter interferometry} 


\author[adleb1,adleb2]{Leonid~I.~Gurvits}
\author[adleb1]{Zsolt~Paragi}
\author[adleb8]{Ricardo~I.~Amils}
\author[adleb1]{Ilse~van~Bemmel}
\author[adleb1]{Paul~Boven}
\author[adleb4]{Viviana~Casasola} 
\author[adleb5]{John~Conway}
\author[adleb45,adleb46]{Jordy~Davelaar}
\author[adleb8]{M.~Carmen~D\a'iez-Gonz\a'alez}
\author[adleb6]{Heino~Falcke}
\author[adleb10]{Rob~Fender}
\author[adleb11,adleb12]{S\'{a}ndor~Frey} 
\author[adleb13]{Christian~M.~Fromm}
\author[adleb8]{Juan~D.~Gallego-Puyol}
\author[adleb8]{Cristina~Garc\a'ia-Mir\a'o}
\author[adleb7,adleb9]{Michael~A.~Garrett}
\author[adleb4]{Marcello~Giroletti}
\author[adleb47,adleb48]{Ciriaco~Goddi}
\author[adleb14]{Jos\'e~L.~G\'omez}
\author[adleb6]{Jeffrey~van~der~Gucht}
\author[adleb15]{Jos\a'e~Carlos~Guirado}
\author[adleb45]{Zolt\a'an~Haiman}
\author[adleb18]{Frank~Helmich}
\author[adleb19]{Ben~Hudson}
\author[adleb20]{Elizabeth~Humphreys}
\author[adleb9]{Violette~Impellizzeri}
\author[adleb3]{Michael~Janssen}
\author[adleb24,adleb85]{Michael~D.~Johnson}
\author[adleb21,adleb22,adleb3]{Yuri~Y.~Kovalev}
\author[adleb3]{Michael~Kramer}
\author[adleb5]{Michael~Lindqvist} 
\author[adleb23]{Hendrik~Linz}
\author[adleb4]{Elisabetta~Liuzzo}
\author[adleb3,adleb22]{Andrei~P.~Lobanov} 
\author[adleb8]{Isaac~L\a'opez-Fern\a'andez}
\author[adleb8]{Inmaculada~Malo-G\a'omez}
\author[adleb30]{Kunal~Masania}
\author[adleb13]{Yosuke~Mizuno}
\author[adleb21,adleb22]{Alexander~V.~Plavin}
\author[adleb31]{Raj~T.~Rajan}
\author[adleb13]{Luciano~Rezzolla}
\author[adleb24,adleb85,adleb6]{Freek~Roelofs}
\author[adleb3]{Eduardo~Ros}
\author[adleb4]{Kazi~L.\,J.~Rygl}
\author[adleb25,adleb26,adleb3]{Tuomas~Savolainen}
\author[adleb27]{Karl~Schuster}
\author[adleb4]{Tiziana~Venturi}
\author[adleb1]{Marjolein~Verkouter}
\author[adleb8]{Pablo~de~Vicente}
\author[adleb2]{Pieter~N.A.M.~Visser}
\author[adleb28]{Martina~C.~Wiedner}
\author[adleb3]{Maciek~Wielgus}
\author[adleb29]{Kaj~Wiik}
\author[adleb3]{J.~Anton~Zensus}

\address[adleb1]{Joint Institute for VLBI ERIC, Dwingeloo, The Netherlands}
\address[adleb2]{Faculty of Aerospace Engineering, Delft University of Technology, Delft, The Netherlands}
\address[adleb8]{Yebes Observatory, IGN, Spain}
\address[adleb4]{INAF Institute of Radio Astronomy, Bologna, Italy}
\address[adleb5]{Department of Space, Earth and Environment, Chalmers University of Technology, Onsala Space Observatory, 439 92 Onsala, Sweden}
\address[adleb45]{Department of Astronomy, Columbia University, New York, NY 10027, USA}
\address[adleb46]{Center for Computational Astrophysics, Flatiron Institute, 162 Fifth Avenue, New York, NY 10010, USA}
\address[adleb6]{Department of Astrophysics, Institute for Mathematics, Astrophysics and Particle Physics (IMAPP), Radboud University, P.O. Box 9010, 6500 GL, Nijmegen, The Netherlands}
\address[adleb10]{Oxford University, UK}
\address[adleb11]{Konkoly Observatory, Res. Centre for Astronomy and Earth Sciences, Budapest, Hungary}
\address[adleb12]{Institute of Physics, ELTE E\a"otv\a"os Lor\a'and University, Budapest, Hungary}
\address[adleb13]{Goethe-Universit\"at Frankfurt, Germany}
\address[adleb7]{Jodrell Bank Centre for Astrophysics, The University of Manchester, UK}
\address[adleb9]{Leiden Observatory, Leiden University, The Netherlands}
\address[adleb47]{Dipartimento di Fisica, Università degli Studi di Cagliari, Monserrato, Italy}
\address[adleb48]{INAF -- Osservatorio Astronomico di Cagliari, Selargius, Italy}
\address[adleb14]{Instituto de Astrof\a'isica de Andaluc\a'ia -- CSIC, Granada, Spain}
\address[adleb15]{Observatorio Astron\a'{o}mico, University of Valencia, Spain}
\address[adleb18]{Kapteyn Institute, University of Groningen, The Netherlands}
\address[adleb19]{KISPE Space Systems Limited, UK}
\address[adleb20]{ESO, Garching, Germany}
\address[adleb3]{Max-Planck-Institut f\"ur Radioastronomie, Auf dem H\"ugel 69, D-53121 Bonn, Germany}
\address[adleb24]{Center for Astrophysics, Harvard \& Smithsonian, Cambridge, MA 02138, USA}
\address[adleb85]{Black Hole Initiative at Harvard University, 20 Garden
Street, Cambridge, MA 02138, USA}
\address[adleb21]{Astro Space Center, Lebedev Physical Institute, Moscow, Russia}
\address[adleb22]{Moscow Institute of Physics and Technology, Dolgoprudnyi, Russia}
\address[adleb23]{Max Planck Institute for Astronomy, Heidelberg, Germany}
\address[adleb30]{Shaping Matter Lab, Faculty of Aerospace Engineering, Delft University of Technology, The Netherlands}
\address[adleb31]{Faculty of Electrical Engineering, Mathematics and Computer Science, Delft University of Technology, Delft, The Netherlands}
\address[adleb25]{Aalto University Department of Electronics and Nanoengineering, Finland}
\address[adleb26]{Aalto University Mets\a"ahovi Radio Observatory, Finland}
\address[adleb27]{IRAM, Grenoble, France}
\address[adleb28]{Observatoire de Paris, PSL University, Sorbonne Université, CNRS, LERMA, Paris, France}
\address[adleb29]{University of Turku, Department of Physics and Astronomy, Tuorla Observatory, Finland}

\begin{abstract} 
Ultra-high angular resolution in astronomy has always been an important vehicle for making fundamental discoveries. Recent results in direct imaging of the vicinity of the supermassive black hole in the nucleus of the radio galaxy M87 by the millimeter VLBI system Event Horizon Telescope and various pioneering results of the Space VLBI mission RadioAstron provided new momentum in high angular resolution astrophysics. In both mentioned cases, the angular resolution reached the values of about 10$-$20~microrcseconds (0.05$-$0.1~nanoradian). Further developments toward at least an order of magnitude ``sharper'' values, at the level of 1~microarcsecond are dictated by the needs of advanced astrophysical studies. The paper emphasis that these higher values can only be achieved by placing millimeter and submillimeter wavelength interferometric systems in space. A concept of such the system, called Terahertz Exploration and Zooming-in for Astrophysics, has been proposed in the framework of the ESA Call for White Papers for the Voayage 2050 long term plan in 2019. In the current paper we present new science objectives for such the concept based on recent results in studies of active galactic nuclei and supermassive black holes. We also discuss several approaches for addressing technological challenges of creating a millimeter/sub-millimeter wavelength interferometric system in space. In particular, we consider a novel configuration of a space-borne millimeter/sub-millimeter antenna which might resolve several bottlenecks in creating large precise mechanical structures. The paper also presents an overview of prospective space-qualified technologies of low-noise analogue front-end instrumentation for millimeter/sub-millimeter telescopes. Data handling and processing instrumentation is another key technological component of a sub-millimeter Space VLBI system. Requirements and possible implementation options for this instrumentation are described as an extrapolation of the current state-of-the-art Earth-based VLBI data transport and processing instrumentation. The paper also briefly discusses approaches to the interferometric baseline state vector determination and synchronisation and heterodyning system. The technology-oriented sections of the paper do not aim at presenting a complete set of technological solutions for sub-millimeter (terahertz) space-borne interferometers. Rather, in combination with the original ESA Voyage 2050 White Paper, it sharpens the case for the next generation microarcsceond-level imaging instruments and provides starting points for further in-depth technology trade-off studies. 
\end{abstract}


\end{frontmatter}

\bigskip
\smallskip

\textbf{List of acronyms}

\begin{tabbing}
Baikal-GVD***  \= Gigaton Volume [neutrino] Detector   \kill
AAReST		\> Autonomous Assembly of a Reconfigurable Space Telescope \\
ADE			\> Axially Displaced Ellipse   \\
AGN			\> Active Galactic Nuclei    \\
ALMA		\> Atacama Large Millimeter Array   \\
AMiBA		\> Array for Microwave Background Anisotropy \\
ANTARES		\> Astronomy with a Neutrino Telescope and Abyss environmental RESearch project \\
Baikal-GVD	\> Gigaton Volume [neutrino] Detector  \\
BUST        \> Baksan Underground Scintillation Telescope \\
CASA		\> Common Astronomy Software Application package \\
CPU         \> Central processing unit \\
CSO			\> Cryogenic Sapphire Oscillator  \\
EHI			\> Event Horizon Imager   \\
EHT			\> Event Horizon Telescope  \\
ERA         \> European Robotic Arm  \\
ESA			\> European Space Agency  \\
FFT			\> Fast Fourier Transform   \\
FPGA		\> Field-Programmable Gate Array   \\
FRB			\> Fast Radio Burst  \\
GNSS		\> Global Navigation Satellite System  \\
GPU         \> Graphics processing unit \\
GRACE		\> Gravity Recovery and Climate Experiment \\
GRMHD		\> General Relativistic Magnetohydrodynamic \\
HEB			\> Hot Electron Bolometers \\
HEMT		\> High Electron Mobility Transistor  \\
HERO		\> HEterodyne Receiver for the Origins Space Telescope \\
HIFI        \> Herschel Heterodyne Instrument for the Far-Infrared \\
IceCube     \> IceCube Neutrino Observatory \\
IF			\> Intermediate frequencies  \\
ISS			\> International Space Station  \\
JPL			\> Jet Propulsion Laboratory   \\
KBR			\> K-band radio-wave ranging   \\
LADEE		\> Lunar Atmosphere and Dust Environment Explorer  \\
LEO			\> Low Earth Orbit   \\
LISA			\> Laser Interferometer Space Antenna   \\
LNA			\> Low-Noise Amplifier   \\
LRI			\> Laser Ranging Instrument   \\
NAOJ		\> National Astronomical Observatory of Japan  \\
MEO			\> Medium Earth Orbits   \\
MMIC		\> Monolithic Microwave Integrated Circuits  \\
NASA		\> National Aeronautics and Space Administration \\
OVRO		\> Owens Valley Radio Observatory \\
PAF			\> Phased Array Feeds  \\
PIMA		\> VLBI software package  \\
RATAN-600	\> Radio Telescope of Academy of Sciences, 600 m in diameter  \\
RF			\> Radio frequencies  \\
SIS			\> Silicon Integrated Systems \\
SMBH		\> Supermassive Black Hole  \\
SOFIA		\> Stratospheric Observatory for Infrared Astronomy  \\
SPICA		\> Space Infrared Telescope for Cosmology and Astrophysics   \\
TDRSS		\> Tracking and Data Relay Satellite System   \\
THEZA		\> TeraHertz Exploration and Zooming-in for Astrophysics   \\
TRL			\> Technology Readiness Level   \\
VLA			\> Very Large Array    \\
VLBA		\> Very Long Baseline Array   \\
VLBI			\> Very Long Baseline Interferometry  \\
VSOP		\> VLBI Space Observatory Program  \\
YTLA		\> Yuan-Tseh Lee Array  \\
\end{tabbing}


\section{Introduction}
\label{s:intro}

Angular resolution is one of the major parameters which define efficiency of an astronomical observing instrument. For the past half a century, the record in angular resolution firmly belongs to the Very Long Baseline Interferometry (VLBI) technique. In recent years, advances of VLBI technologies and new data processing algorithms enabled the global Event Horizon Telescope (EHT) collaboration to image the black hole shadow in M87* (\citep{eht-paperI} and references therein) and probe the innermost regions of Active Galactic Nuclei (AGN) jets \cite{Kim2020,Janssen2021} at the wavelength of 1.3~mm with an angular resolution of about 20~$\mu$as. The Space VLBI mission RadioAstron operated at wavelengths down to 1.3~cm and baselines up to $\sim$30~Earth diameters thus enabling angular resolution reaching $\sim$10~$\mu$as \cite{Kardashev+2013}. However, these observing capabilities at the level of angular resolution of tens of microarcsconds do not exhaust the science drives toward even sharper radio astronomy `vision', both on the ground and in space. In fact, over the past four decades, there were a number of studies aiming to achieve an even higher angular resolution, including several design studies of Space VLBI missions (\cite{Gurvits-2018IAC,Gurvits-2020} and references therein). 

The range of electromagnetic spectrum between hundreds of gigahertz and several terahertz (hereafter called a terahertz range for brevity; it corresponds to wavelengths from $\sim$1~mm down to sub-millimeters) is used for many diverse scientific and technological applications. In astronomy, spectroscopic studies in the THz range provide information on spectral lines of molecules and atoms that are essential to understanding of astrochemistry of various constituencies of galactic matter and evolution of galaxies, stars and planets. For the VLBI technique, the millimeter domain provides the highest angular resolution accessible by any Earth-based telescopes as demonstrated recently by the EHT. However, THz radiation is absorbed by the water vapour in the Earth’s atmosphere. Therefore, observations at frequencies up to $\sim$350~GHz are practically possible at very special places -- either at high altitudes or in extremely cold and dry regions (e.g., Antarctica), or on stratospheric aircraft or balloons. Thus, a radical solution of the problem of atmosphere opacity is in placing THz telescopes in space.

Angular resolution of a telescope is proportional to the observing wavelength and inversely proportional to its aperture size, which, for an interferometer, is its baseline. Therefore, sharpening the angular resolution in principle can be achieved by either shortening observing wavelengths (as demonstrated by the EHT) or increasing the interferometer baseline (Space VLBI). However, it is the combination of these two approaches that offers the most radical improvement in angular resolution. Among various advantages of a Space VLBI mission operating at THz frequencies we underline two which cannot be achieved in principle by any other means: \\
-- the ability to operate at frequencies higher than those of EHT will offer the unique opportunity to study a new population of sources associated with resolvable black hole shadows which are opaque at 230~GHz (cf. \cite{Pesce+2021}); \\
-- long space baselines at these high frequencies will make possible unique probes of black hole spin and spacetime properties (see subsection~\ref{s:subrings}). 
Angular resolution of Earth-based VLBI systems cannot be significantly higher than that of EHT. Even at the highest operational frequencies permitted by the Earth atmosphere ($\sim$350~GHz, equivalent to the wavelength of $\sim$1~mm), the angular resolution is merely 40$\%$ higher than demonstrated so far, $\sim$20~$\mu$as. Thus, a scientifically transformational leap by an order of magnitude toward a microarcsecond angular resolution can be achieved only by placing a VLBI system in space.

In the wake of the EHT and RadioAstron results, a global initiative aiming to present a case for a Space VLBI mission able to operate at mm and sub-mm wavelengths has taken off the ground, see \cite{Charlottesville-2020} and references therein. The concept of TeraHertz Exploration and Zooming-in for Astrophysics (THEZA) is one of the components of this global initiative. It was developed in 2019 in response to the ESA's Call for White Papers for the long-term plan Voyage~2050 \cite{Voyage-2050-21}. The concept considers a space-borne mm/sub-mm interferometric system able to image celestial radio sources with angular resolution reaching single-digit microarcseconds. The THEZA White Paper \cite{Gurvits+2021} presents the science case and charters briefly engineering challenges of such a mission and possible ways to their resolution. In the current paper we describe new recently developed scientific topics which strengthen the main objectives of the THEZA concept. We also suggest several novel engineering approaches which might make the THEZA mission a reality. 

We see the goal of the current paper in presenting a conceptual study and providing starting points for future in-depth design and engineering trade-off studies of a space-borne terahertz interferometric system. Section~\ref{s:sci-case} describes additional science applications for the THEZA concept which strengthen the case described in the original ESA White Paper \citep{Gurvits+2021}. In particular, they are based on the recent results stimulated by the EHT studies of M87* (e.g., \citep{Johnson+2020,Wielgus-2021ar,Palumbo-Wong-2022} and references therein) and other astrophysical objects (e.g., \citep{Kim2020,Janssen2021}), published after the submission of the Voyage~2050 THEZA White Paper. Section~\ref{s:mi-concpt} describes a set of major benchmark specifications for a mission based on the THEZA concept. The following sections~\ref{s:ant-syst}--\ref{s:t-604} address key mission components which have crucial impact on the overall ability of the system to address science applications presented in \cite{Gurvits+2021} and section~\ref{s:sci-case}. Finally, section~\ref{s:concl} offers brief concluding remarks on the THEZA concept.

\begin{figure}[t]
  \centering
\includegraphics[width=.40\textwidth]{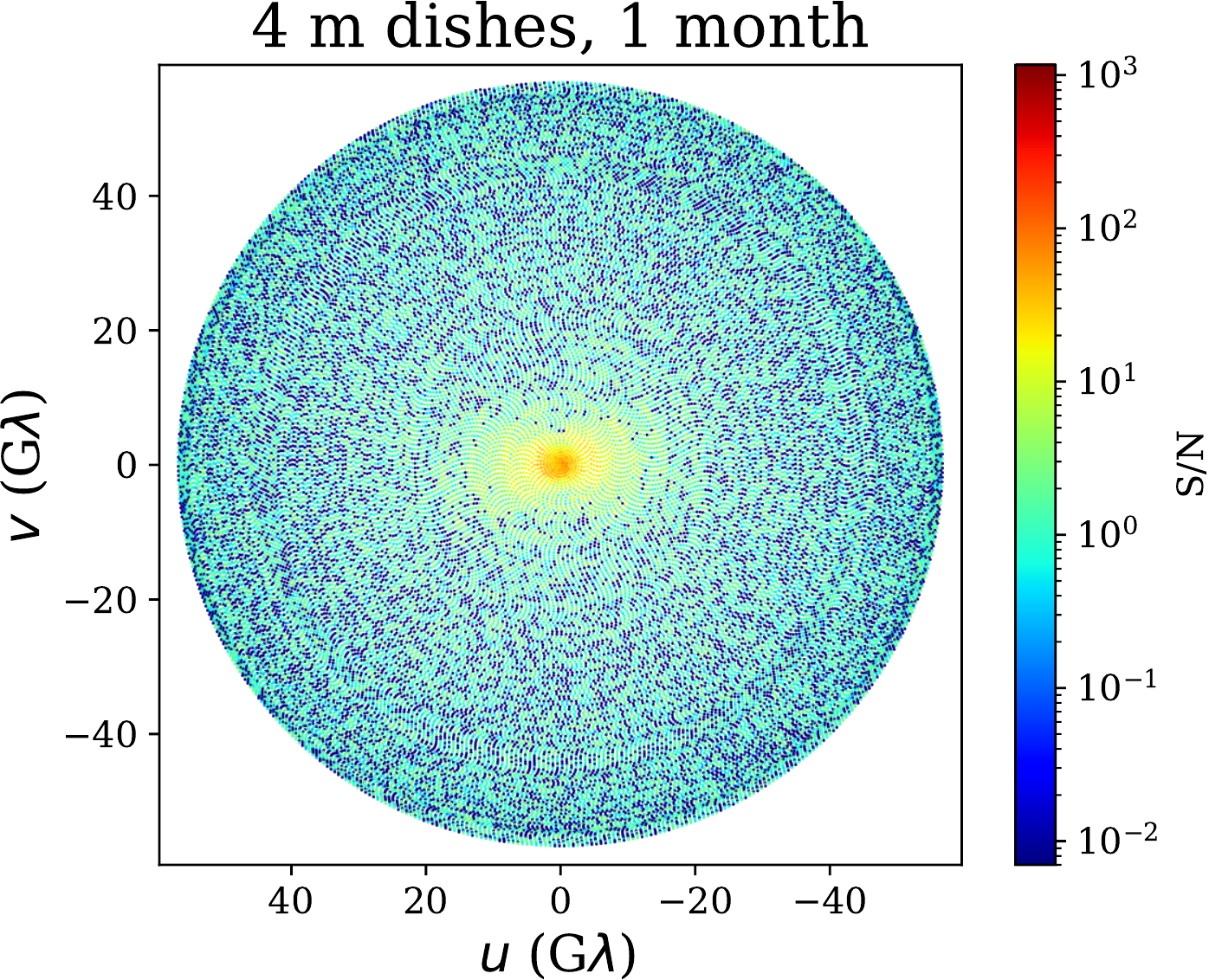}
\includegraphics[width=.40\textwidth]{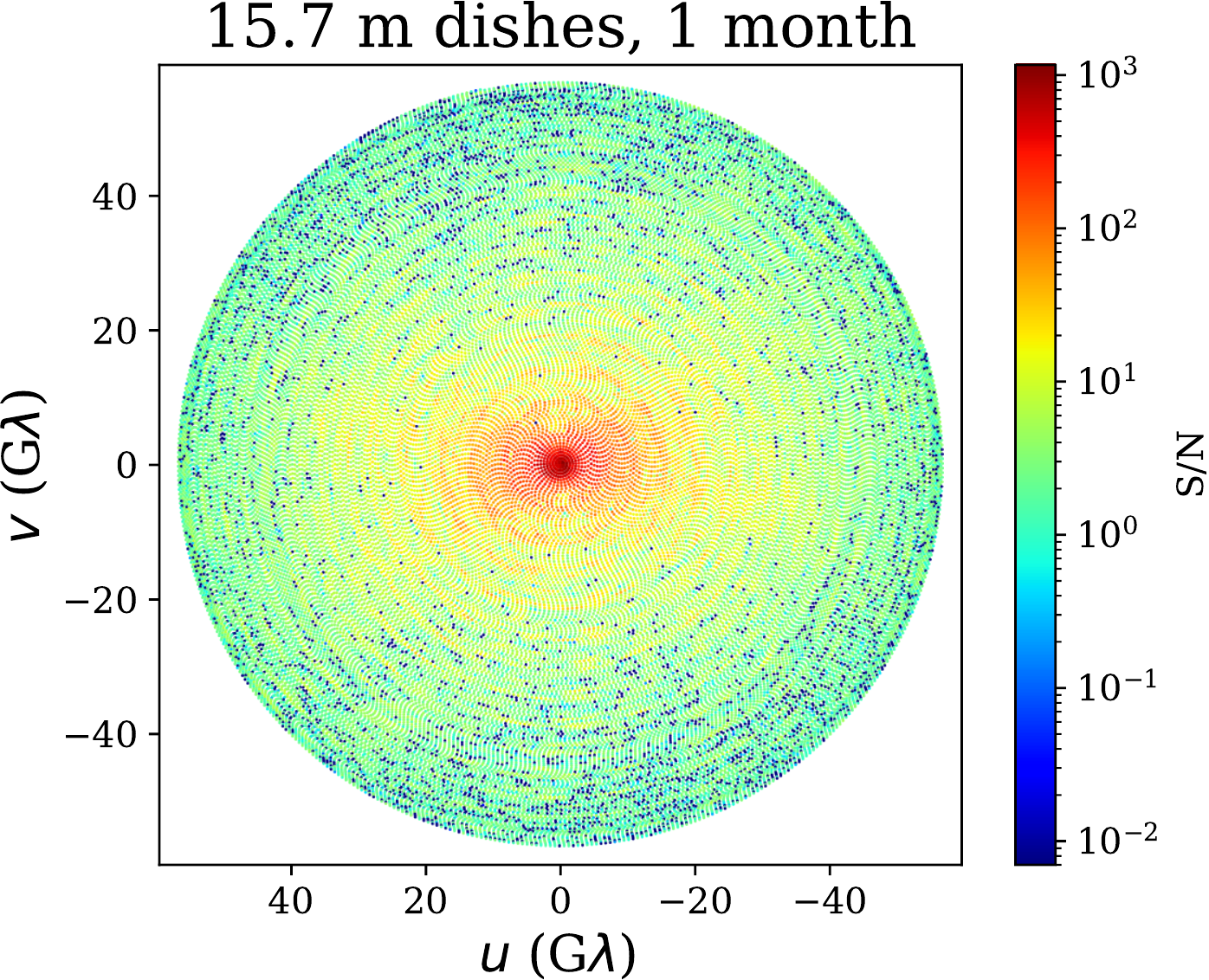}\\
\includegraphics[width=.40\textwidth]{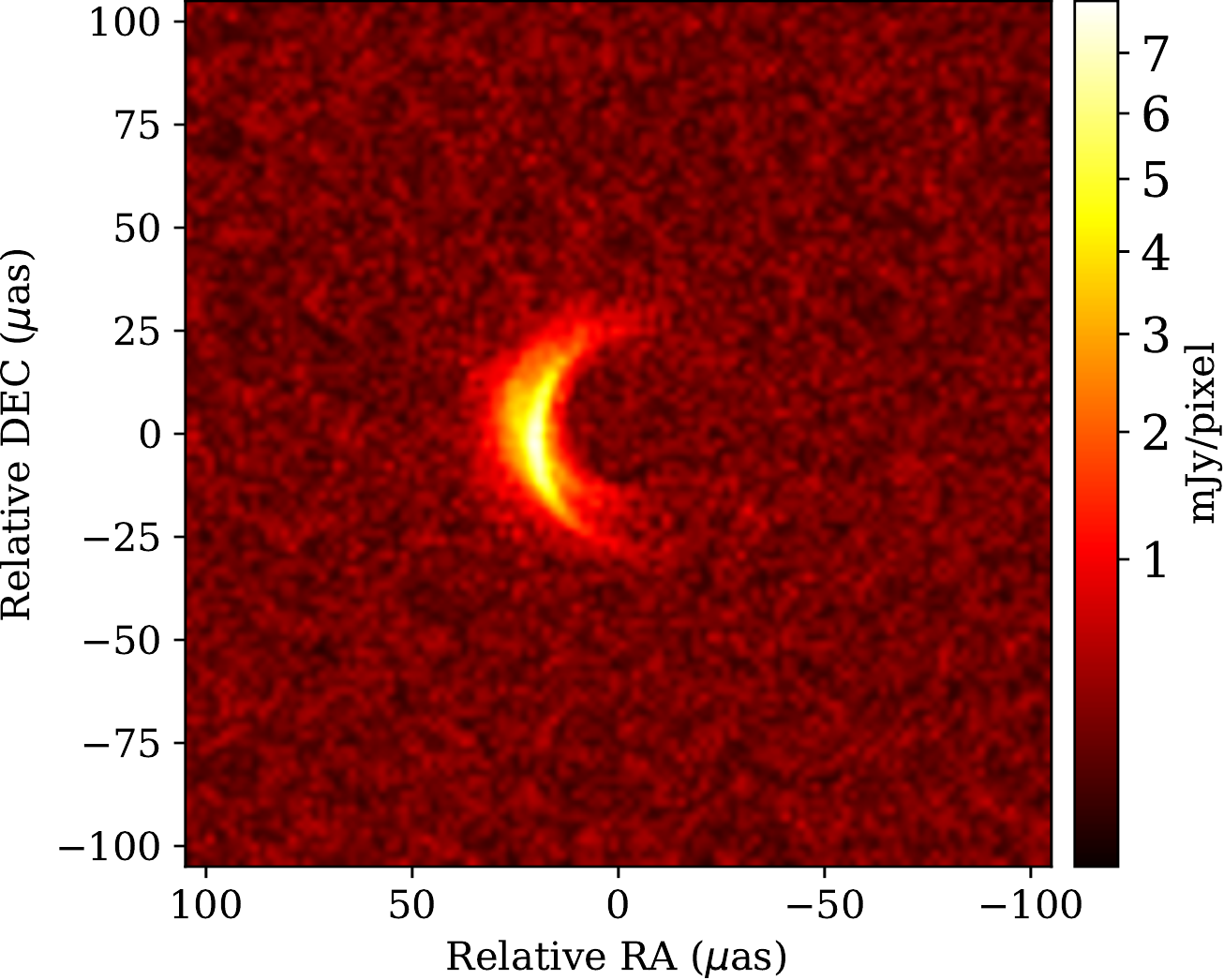}
\includegraphics[width=.40\textwidth]{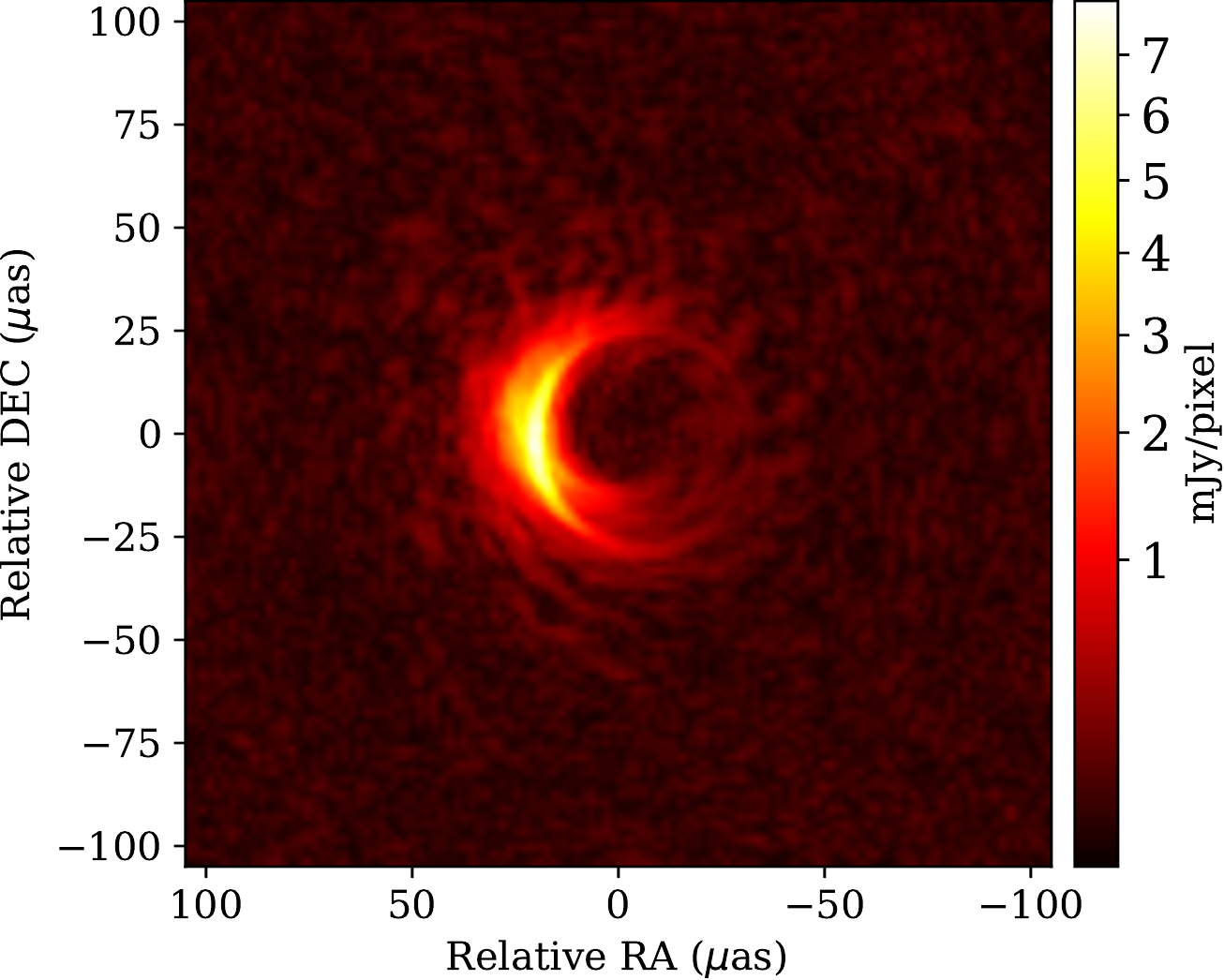}
\caption{Top: Signal-to-noise ratio of individual visibilities (integration times 0.5-7.5 minutes depending on baseline length) when observing a GRMHD model of Sgr A* \citep[model 39 from][]{Moscibrodzka2014} at 690 GHz with the Event Horizon Imager (EHI) system, using system noise parameters from \cite{Roelofs+2019} and three dishes with a diameter of 4 (left) or 15.7 (right) m. Bottom: images reconstructed by taking an FFT of the visibilities on a $116\times116$ pixel grid. The normalized cross-correlation when comparing to the average image of the observed movie is 0.89 and 0.97 for 4 and 15.7-meter dishes, respectively.}
   \label{f:sgrasims}   
\end{figure}

\section{THEZA science case}
\label{s:sci-case}


THEZA is a concept of a multi-purpose astrophysical facility. Its main specifications aim at supporting  transformational studies of supermassive black holes (SMBH) with unprecedented angular resolution and sensitivity, thus enabling investigation of the physics of space-time in the strong-field regime inaccessible by any other experimental technique. However, as demonstrated in the THEZA White Paper \cite{Gurvits+2021}, the concept offers a wide range of science applications ranging from population studies of active galactic nuclei, progenitors of gravitational wave events and other multi-messenger phenomena, formation of stellar and planetary systems, astrochemical studies and search for technosignatures. In this paper, while focusing on several new considerations for the mission architecture and required technologies, we present several new science cases which reflect recent developments and results. We note that many science objectives of THEZA coincide or are highly synergistic to those developed for the Next Generation EHT (ngEHT, \cite{Doeleman-2021AAS,ngEHT-ASTRO2020}).

\subsection{Imaging of supermassive black holes on event horizon scales}

Imaging of SMBH is one of the main science objectives of the THEZA concept. The long baselines and high frequencies attainable from space allow an order-of-magnitude improvement in image resolution and fidelity compared to ground-based observations with the Event Horizon Telescope, which published the first image of a black hole shadow in 2019 \citep{eht-paperI,eht-paperII,eht-paperIII,eht-paperIV,eht-paperV,eht-paperVI}. Sharper event-horizon scale images of supermassive black holes will help test general relativity to high precision, measuring black hole properties such as mass and spin, constrain plasma models of the accretion flow, and provide transformational input into cosmological tests \cite{Goddi+2017}.

A concept of Event Horizon Imager (EHI), a predecessor of discussed here THEZA concept, was analysed and simulated in \citep{Roelofs+2019,Roelofs+2020}. It involved two or three satellites in Medium Earth Orbits (MEO). Due to a slight difference in the orbit radii, the baselines form a dense spiral in the $uv$-plane as the satellites orbit Earth and drift apart, so that information is available on all baseline lengths and directions up to the maximum baseline length, which gives nominal resolution of 3.5 $\mu$as at 690 GHz. For comparison, the current nominal resolution of the EHT at 230 GHz is 23 $\mu$as.

Figure \ref{f:sgrasims} shows how such a system is limited by the attainable signal-to-noise ratio (S/N) rather than the $uv$-coverage. An EHI observation of a general relativistic magnetohydrodynamics (GRMHD) simulation of Sgr A* \citep{Moscibrodzka2014} was simulated following \cite{Roelofs+2019}, using different antenna dish sizes. With three 4-meter dishes, a sharp crescent feature can be seen after integrating for one month, but with 15.7-meter dishes (the equivalent of 19 4-meter dishes assuming a phasing efficiency of 0.9, see also section \ref{s:ant-syst}), the full photon ring can be distinguished clearly and with high precision, as high-S/N visibilities are detected up to baseline lengths of several tens of G$\lambda$. The difference in image quality is attested by the normalized cross-correlation \citep[nxcorr, e.g.][]{eht-paperIV}, which is 0.89 and 0.97 for 4 and 15.7-meter dishes, respectively, when comparing to the average image of the observed GRMHD movie. Such a sharp image of the photon ring can be used for precise tests of general relativity, which predicts its size and shape. Furthermore, it can be used to put strong constraints on the black hole mass and spin, and plasma parameters \citep{Gucht2020, Roelofs2021}. 

\subsection{Black hole's photon rings}
\label{s:subrings}

\begin{figure}[th!]
  \centering
\includegraphics[width=.80\textwidth,trim={0cm 1cm 0 0},clip]{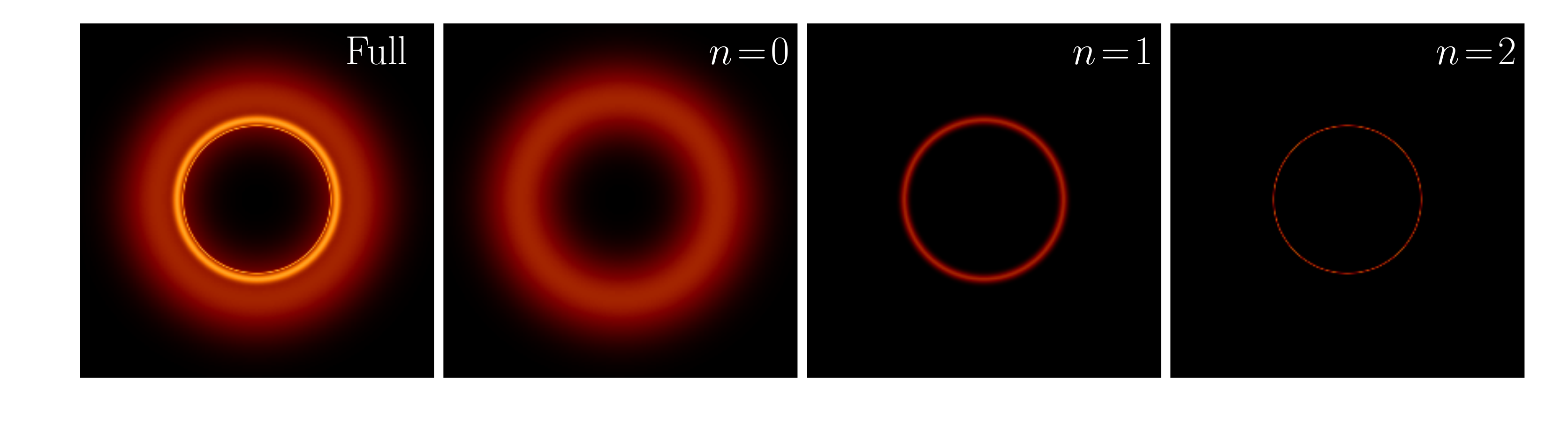}
\caption{
A toy model of an image of a black hole viewed face-on, with its decomposition into the direct image ($n=0$) and first two photon rings ($n=1,2$) shown. The photon rings have approximately constant brightness but become exponentially narrower with increasing index $n$. They converge to a theoretical critical curve, which has a size and shape that depend exclusively on the spacetime geometry.
}
   \label{f:photon_ring}   
\end{figure}

\begin{figure}[th!]
  \centering
\includegraphics[width=.9\textwidth]{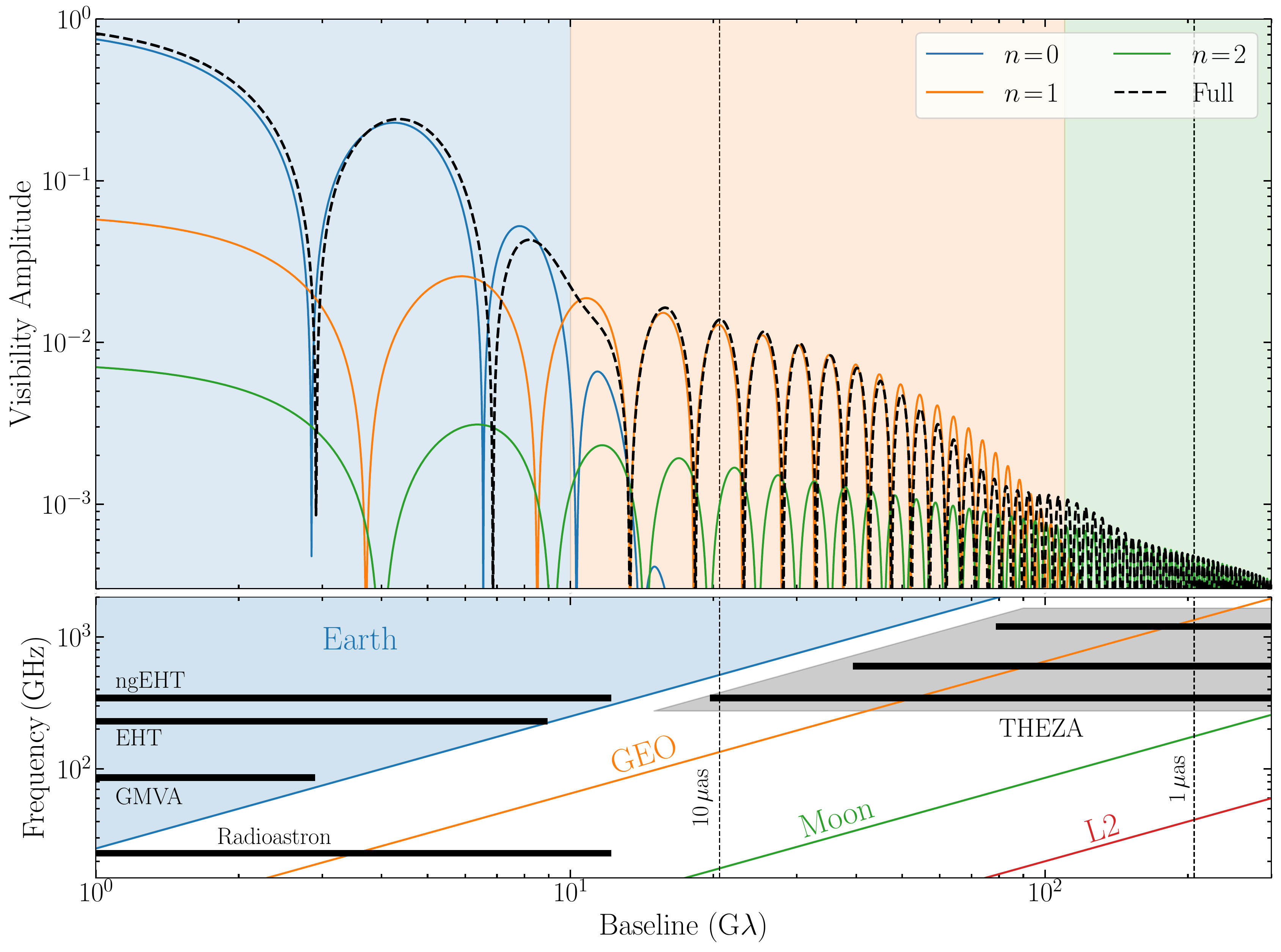}
\caption{
Top: Fourier domain structure of the black hole images shown in Figure \ref{f:photon_ring}. The photon rings are expected to dominate the total signal at very long baselines. Bottom: capabilities of the highest-resolution existing instruments. The limitation of the ground-based instruments is shown as a blue-shaded area. A resolving power of Earth - geostationary orbit (GEO), Earth-Moon, and Earth-L2 orbits is indicated. THEZA would enable transformational ultra-high resolution science of black hole photon rings and entirely new tests of general relativity. 
}
   \label{f:photon_ring_vis}   
\end{figure}

At the currently highest resolution, baselines up to 10~G$\lambda$ corresponding to a few Schwarzschild radii for the objects Sgr~A$^{*}$ and M87$^{*}$, it is necessary to make strong assumptions on the accretion flow properties in order to constrain the spacetime geometry \cite{eht-paperV,eht-paperVI,Gralla+2019}. Indeed, current EHT measurements do not even significantly constrain the black hole spin of M87$^{*}$. This weakens the robustness of the spacetime tests, as our understanding of the detailed physics of astrophysical plasma, relevant for the image formation, remains largely incomplete (e.g., \cite{Vincent+2021,Gralla-2021}). However, owing to the extreme lensing near the photon shell, images of black holes contain a sequence of demagnified copies of the direct image -- photon rings \cite{Johnson+2020,Gralla+Lupsasca-2020}, as shown in the Figure \ref{f:photon_ring}. The photon rings are less sensitive to the accretion flow properties than the direct image, observed by the EHT \cite{eht-paperI}. Moreover, sharp image domain features decay slowly in the visibility (Fourier) domain, dominating the signal at high spatial frequencies (see the upper panel of Figure \ref{f:photon_ring_vis}), thus rendering space radio interferometry a well-suited tool to probe the photon ring structure of the black hole images \cite{Johnson+2020,Vincent+2021}. However, constraining photon ring structure in the VLBI data requires probing significantly higher spatial frequencies than has been enabled by the EHT or RadioAstron, as indicated in the bottom panel of Figure \ref{f:photon_ring_vis}.

Tests of general relativity involving the second order photon ring require baseline lengths exceeding ${\sim}100\,{\rm G}\lambda$ for M87$^{*}$ or Sgr~A$^{*}$ \cite{Johnson+2020,Gralla++2020}. While such an extreme spatial resolution may necessitate a system operating at the EHT frequencies with the physical baselines comparable with those of RadioAstron, spacetime geometry tests involving the first order photon ring have also been proposed, allowing to probe the black hole spin \cite{Broderick+2021ar} or spacetime deviation from the Kerr solution \cite{Wielgus-2021ar}. The first order photon ring analysis requires spatial frequencies at most several tens of giga-wavelengths, as shown in Figure \ref{f:photon_ring_vis}. What is more, the other factor limiting feasibility of the photon ring observations is related to the source-intrinsic opacity, which reduces the prominence of the strongly sensed features and ultimately truncates the maximal order of observable photon rings. These spurious effects of opacity are reduced with the increasing observing frequency, rendering high-frequency mission concepts favourable for this kind of observations. The proposed capabilities of THEZA would enable observations of both first and second ($n=1,2$) photon rings.

\subsection{Supermassive black holes and relativistic outflows}
\label{S:VLBP}

\begin{figure*}[t]
\centering
\includegraphics[width=0.85\textwidth]{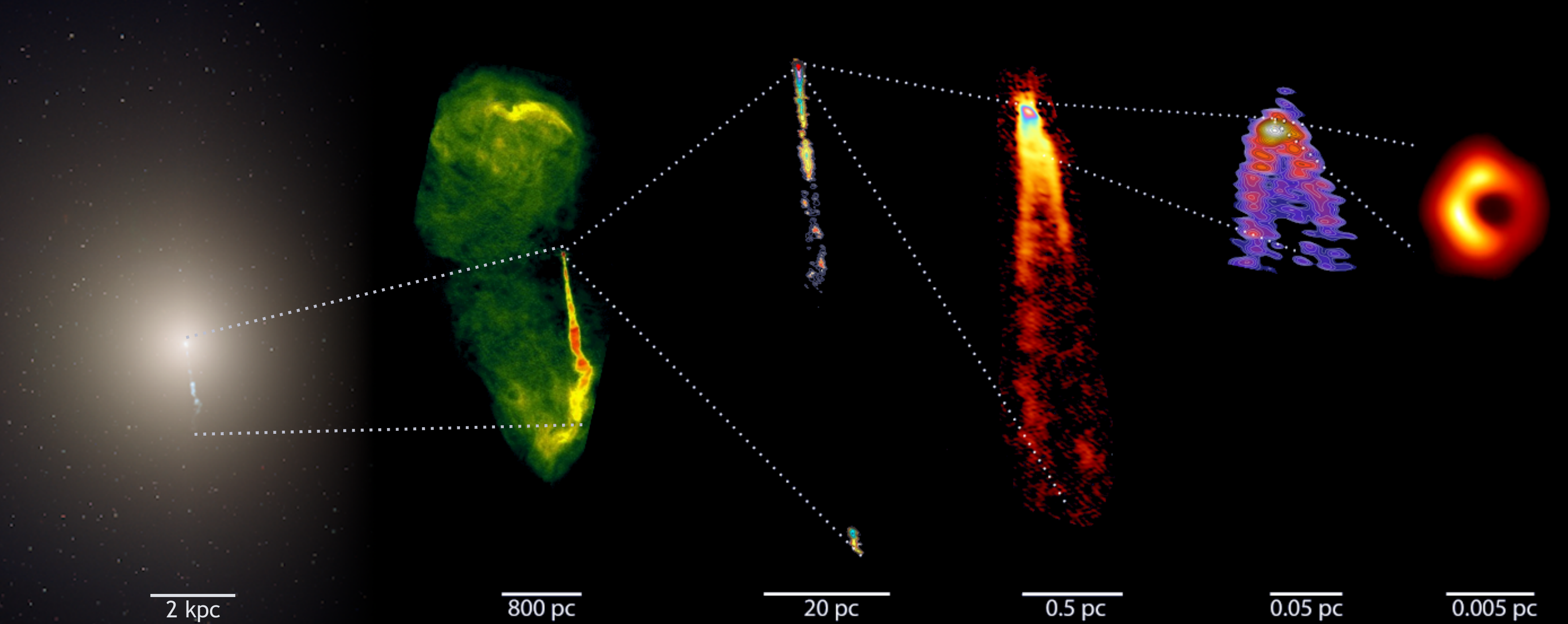}
\caption{M87 hosts the supermassive black hole captured by the Event Horizon Telescope in the first-ever image of a black hole. From left to right: Hubble optical image of the M87 field; VLA radio continuum image at the wavelength 20\,cm; VLBA at 20\,cm; VLBA at 7\,mm; GMVA at 3\,mm; and EHT image at 1\,mm (the images are all rotated by 90$^\circ$. The figure is adapted from \cite{2019ARA&A..57..467B} (with references therein for the shown five left images) and \cite{eht-paperI} for the right-most EHT image. The central supermassive black hole shown at the right is responsible for the enormous energy driving the relativistic jet, shown at all scales in the figure.}
\label{fig:m87_jet-bh}
\end{figure*}

The extremely high angular resolution offered by the VLBI technique enables us to study cores of active galactic nuclei at centimeter wavelengths with ``sharpness'' a thousand times better than regular optical observations. In particular, it makes possible investigation of mass accretion onto supermassive black holes and formation of an accretion disk that surrounds the event horizon. In the case of the most powerful active galactic nuclei, known as blazars (i.e, active galactic nuclei with jets pointed towards us), magnetic fields either anchored in the innermost accretion disk \citep{1982Blandford} or black hole ergosphere \citep{1977Blandford} extract some of this material forming a pair of very powerful and highly collimated relativistic jets that extend far beyond the nucleus area of the host galaxy (see Fig.~\ref{fig:m87_jet-bh} for the archetypal jet in M87). Relativistic electrons in the jet, threaded by large-scale magnetic fields, radiate most of their energy as synchrotron and perhaps inverse Compton emission across the entire spectrum, from radio to $\gamma$-rays.

While several studies over the past decades have provided a broad picture of physical processes like accretion onto SMBHs, ignition of  galactic nuclei and propagation of relativistic jets, the currently achievable angular resolution  with ground-based VLBI is insufficient to resolve  all these processes in the SMBH vicinity -- i.e., on scales crucial to test competing theoretical models. Two recent major instrumental improvements have partially overcome this limitation, providing a first glimpse into the inner works of SMBHs. On one hand, the Space VLBI mission RadioAstron has allowed to increase the virtual size of our VLBI telescopes to as large as the distance to the Moon, achieving angular resolutions as small as $\sim$10 microarcseconds. On the other hand, the participation of ALMA as a phased array in VLBI observations at millimeter wavelengths \cite{Matthews+2018,Goddi+2019} has allowed the EHT to reconstruct images with similar angular resolutions, but at higher frequencies and therefore lifting the opacity curtain that affects longer wavelengths, providing the first clear look into a black hole \citep{eht-paperI}.

The RadioAstron Space VLBI mission \citep{Kardashev+2013} featured a 10\,m radio telescope on board of the \textit{Spektr-R} satellite. With an apogee of $\sim$350 000\,km, Space VLBI observations with RadioAstron made possible imaging of blazar jets in total and linearly polarised intensity with an unprecedented resolution at the wavelength of 1.3\,cm \citep[e.g.,][]{Gomez+2016}. Three Key Science Programmes (KSPs) on AGN imaging have collected data since 2013 to study the launching, collimation, and magnetic field properties of AGN jets, while the AGN survey studied the brightness temperature of their cores \citep[e.g.,][]{2020AdSpR..65..705K}.

First RadioAstron polarimetric observations at the shortest operating wavelength of 1.3\,cm targeted BL\,Lac. Earth--Space fringes were detected up to a maximum projected baseline length of about 8 Earth diameter, allowing to image the innermost jet in total and linearly polarised intensity with an unprecedented angular resolution of 21\,$\mu$as \citep{Gomez+2016}. Gradients in Faraday rotation and intrinsic polarization vectors were found as a function of position angle with respect to the VLBI core, suggesting that the jet launching region in BL\,Lac is threaded by large-scale helical magnetic fields, as expected from theoretical models. Similarly, intra-day variable blazar 0716$+$714 was observed with RadioAstron at 1.3\,cm probing the vicinity of the central black hole with an angular resolution of 24\,$\mu$as \citep{Kravchenko_2020}. High linear polarisation was detected in a very compact jet region 19\,$\mu$as in size located at about 60\,$\mu$as from the central engine. The VLBI core was resolved out into a highly bent structure, which suggests that the jet viewing angle lies inside the opening angle of the jet, which in turn may explain the intra-day variability that characterises this source through rapid variable Doppler boosting. 

While RadioAstron is allowing us to probe the magnetic field in the innermost jet regions of the brightest blazars (see also \cite{2015A&A...583A.100L,2021A&A...648A..82P,2017A&A...604A.111B}), the EHT has recently provided the first image of linearly-polarised emission at the event horizon scale in M87$^{*}$, encoding the magnetic field structure present in the system \citep{2021ApJ...910L..12E,2021ApJ...910L..13E,Goddi+2021}. The images show that the southwest part of the ring is highly linearly polarized, with polarization position angles arranged in a nearly azimuthal pattern that may result from organized poloidal magnetic fields in the vicinity of the central black hole. Comparison with GRMHD models suggest that the near-horizon magnetic fields are dynamically important, with strengths of the order of $1-30$~G, consistent with a magnetically arrested accretion disk.

The natural step forward in our quest for the sharpest astronomical images would be to combine the short wavelength observations of the EHT, and the long baselines of RadioAstron into a sub-mm space VLBI interferometer that for the first time will be capable of addressing the fundamental questions of how gravity works in the strong-field regime near the event horizon, how accretion leads to the formation of jets, and how the latter propagate to drive galaxy evolution at cosmological scales.


\subsection{Central engines of AGN: the case of Cen~A and 3C~279}
\label{ss:cen-A}

Next to horizon-scale science on Sgr\,A* and M87*, the EHT also studies the jets launched by AGN for sources where the black hole shadow cannot be resolved.
Thanks to the resolving power and high observing frequency of the EHT, such observations probe the launching region of extragalactic radio jets close to the black hole \cite{2017Boccardi}.
Due to the well known jet core-shift effect \cite{1998Lobanov}, these jet launching regions are synchrotron-self-absorbed and therefore unobservable optically thick regions for lower frequency observations.
Going towards higher resolution and observing frequencies with THEZA will allow us to zoom in further into the heart of these AGN. Thereby it will be possible to study the jets in greater detail, in particular their transverse structure close to the black hole, which will inform us about how these jets are launched -- by the black hole magnetosphere \citep{1977Blandford} or the accretion flow \citep{1982Blandford}.
And in many of such sources, we expect to be able to image the black hole shadow together with the footprint of the jet \citep{Pesce+2021}.
This will enable survey studies of black hole images and resolve the symbiotic connection between black hole accretion flows and jet formation \citep{1995Falcke}, which underlies the fundamental plane of black hole activity from stellar-mass to supermassive black holes \citep{2003Merloni, 2004Falcke}.

In particular, EHT images of the jets in Centaurus~A (Cen~A) \cite{Janssen2021} and 3C\,279 \cite{Kim2020} have recently been published.
Cen~A is the closest radio-loud AGN to Earth and therefore an ideal VLBI target for instruments that are able to observe at a declination of \ang{-43}. With a 25\,$\mu$as resolution, the EHT has imaged a strongly edge-brightened jet on sub-light-day scales.
Based on an observed core-shift and fundamental plane relations \citep{2003Merloni, 2004Falcke}, it was found that space VLBI observations at THz frequencies are required to peer down towards the event horizon such that the black hole shadow in this source can be resolved in a bright and optically thin emission region. At such high observing frequencies, a baseline length of $\sim 8000$\,km would yield the required resolving power of $1.4\,\mu$as.
Incidentally, this resolution combined with the proximity of Cen~A would enable us to study an extragalactic radio source on scales of a few astronomical units for the first time.

3C\,279 is an archetypal blazar source with a black hole mass of $\sim10^9\,M_\odot$ at a distance of 5\,Gly from Earth.
In the EHT observations, the jet core appears to be elongated perpendicular to the jet direction.
With higher resolution and higher frequency measurements, the nature of this component can be revealed.
It could be the base of an edge-brightened jet, it will be possible to distinguish between the two outer jet arms with a higher spatial resolution offered by space VLBI.
Similarly, a spatially bent jet can be identified with higher resolution measurements.
The possibilities of a standing shock or enormous accretion disk can be discriminated with information about the component's spectral index, between the EHT's 230\,GHz band and THz frequencies.

\subsection{Blazars as sources of TeV--PeV neutrinos}

Astrophysical neutrinos of TeV energies have been convincingly detected by the IceCube experiment since 2012 \citep{IceCubeFirst26}. Indications of the astrophysical high-energy neutrino flux were also found by the ANTARES and Baikal-GVD experiments. Despite all these observations, the origin of high energy astrophysical neutrinos remained unknown until recently. Blazars have been considered a possible class of neutrino sources since the early days of multi-messenger astronomy. However, no significant connection between neutrino events and $\gamma$-ray loud blazars has been found. This contrasts with the association of a single high-energy neutrino event with a $\gamma$-ray flare in the blazar TXS~0506+056 and an excess of low-energy events from the same direction \citep{icecubecollaborationNeutrinoEmissionDirection2018}.

\begin{figure*}[t]
    \centering
\includegraphics[width=0.74\textwidth]{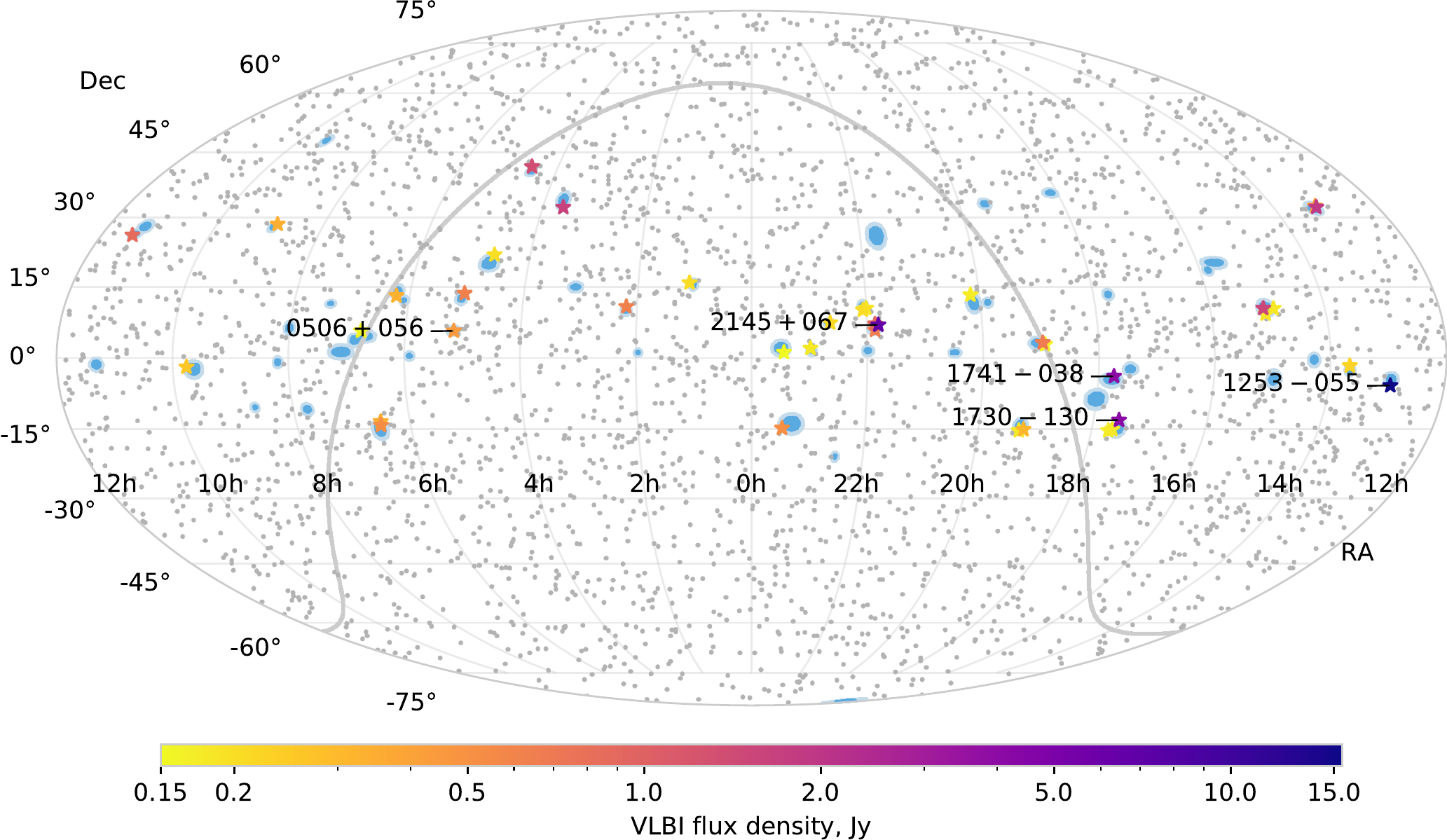}
\caption{
IceCube event locations on the sky, represented by blue ellipses. Stars represent all blazars from a complete VLBI sample within neutrino error regions. Other members of the VLBI sample are shown by gray dots.
The labeled objects denote four blazars with the strongest parsec-scales jets that are the most probable neutrino associations according to our analysis; we also show the location of the first neutrino association TXS~0506+056.
See details in \citep{neutradio1}.}
\label{f:skymap}
\end{figure*}

Recently, it was established \citep{neutradio1,neutradio2} that astrophysical neutrinos with energies from TeV to PeV are produced by bright blazars.  
Figure~\ref{f:skymap} presents positions of IceCube neutrinos above 200~TeV and radio-bright blazars. 
Comparison of the VLBI-selected blazar sample with IceCube neutrino tracks yields the post-trial significance of the directional association of $4.1\sigma$; the probability of a chance coincidence is $p=4\cdot10^{-5}$. There are more than 70 VLBI-selected radio-bright blazars that emit neutrinos of these energies. Moreover, these authors have found a temporal correlation of high-energy neutrino arrivals with radio flares at frequencies above 10~GHz observed by the \mbox{RATAN-600} telescope \citep{neutradio1,2020AdSpR..65..745K}. The most pronounced example is PKS~1502+106 that experienced a major flare in 2019 \citep{r:kielmann1502ATel}. These associations are confirmed based on OVRO and Mets\"ahovi radio monitoring analysis \citep{2021A&A...650A..83H}. We also note a detection of three neutrinos of different energies in December 2021 from the direction of the BL~Lac object PKS~0735$+$17 by IceCube, Baikal-GVD and BUST (GCN\footnote{https://gcn.gsfc.nasa.gov, accessed 2022.01.09} \#31191, ATels\footnote{https://astronomerstelegram.org, accessed 2022.01.09} \#15112, \#15143). The coincident electro-magnetic flare  was reported in gamma-rays (ATels \#15099, \#15129), X-rays (ATels \#15102, \#15108, \#15109, \#15130), optical band (ATels \#15098, \#15100, \#15136) and radio band (ATel \#15105). 

Both the association with VLBI-selected blazars and temporal correlation on scales of months and years indicate that neutrinos are emitted from central parsec-scale regions of active galactic nuclei. Their emission occurs predominantly along the jet direction due to beaming effects. Ultra-relativistic protons up to $10^{16}$~eV and X-ray photons are required to produce observed neutrinos. Estimates indicate that there are enough blazars in the sky to explain the majority of the astrophysical neutrino flux derived by IceCube. Radio-bright blazars associated with neutrino detections have very diverse $\gamma$-ray properties. This suggests that $\gamma$-rays and neutrinos may be produced in different regions of blazars and might not be directly related. A narrow jet viewing angle is, however, required to detect either neutrinos or electromagnetic emission.

The intriguing question which remains open is how protons are accelerated to relativistic energies, as well as details of the neutrino production mechanism. Is it p+p or p+$\gamma$? If the latter, how are the photons produced, how far does it happen from the central engine? Is neutrino production accompanied by new ejecta of relativistic plasma, possibly by shocks which accelerate protons? What is the cause, and what is the effect?
A high angular resolution at the $\mu$as scale and low synchrotron opacity are needed to address these questions and study the region around SMBH and in the jet origin.
Both will be provided by THEZA imaging observations within the proposed specifications with high dynamic range and fidelity.

\section{THEZA mission concept}
\label{s:mi-concpt}

Science objectives of the THEZA concept are presented in the original ESA Voyage 2050 White Paper \cite{Gurvits+2021} and amended with several new cases described in section \ref{s:sci-case} above. Summarily, they all focus on astrophysical studies requiring angular resolution exceeding by at least an order of magnitude the 'sharpest' results achieved to date with the Earth-based EHT \cite{eht-paperI} and Space VLBI mission RadioAstron \cite{Gomez+2021}. Arguably, the most innovative and challenging in terms of interferometric applications in astrophysics is the science case of photon rings described in subsection \ref{s:subrings}. As shown in the bottom panle of Fig.~\ref{f:photon_ring_vis}, the angular resolution required by photon ring studies is in the range of single-digit  microracseconds. We therefore adopt a value of angular resolution of 1~$\mu$as as the benchmark goal for the THEZA concept. Furthermore, as is clear from the same Fig.~\ref{f:photon_ring_vis}, this angular resolution would require observations at frequencies exceeding those of EHT, i.e. what we call in this paper terahertz frequencies. 

The next major THEZA specification parameter is the interferometric baseline sensitivity. The well substantiated recent study \cite{Pesce+Palumbo+2021} provides an estimate of the number $N(\vartheta,S_{\nu})$ of observable SMBH shadows and photon rings as a function of the angular size $\vartheta$ and frequency-dependent flux density $S_{\nu}$ (we reproduce here the symbolic denominations of these values in the cited paper for the reader's convenience). As shown in this study (see Fig.~11 and Table~1 in \cite{Pesce+Palumbo+2021}), there is a strong interdependence between the required angular resolution and interferometric baseline sensitivity for a given number of target sources. The authors conclude that at the single-digit microarcsecond resolution, achieving a 100-fold increase in the number of observable sources at 230~GHz with a baseline consisting of a 10-m telescope and phased ALMA would require the time-bandwidth product of the order of $\sim 3 \times 10^{14}$ (c.f., Table~2 in \cite{Pesce+Palumbo+2021}). Such the value would necessitate several hours of on-source integration and bandwidth of no less than 32~GHz. At present, these benchmark values are beyond reach technologically for both, the ground-based facilities at the highest frequency bands of the atmosphere transparency, and spaceborne systems for higher observing frequencies. However, these current technology limitations are not of fundamental physical nature.  Thus we adopt them as a set of starting reference parameters for investigation of their improvement in further THEZA studies. The baseline sensitivity is defined by the antennas' aperture (diameter), noise characteristics of their receivers and the time-bandwidth product. The following sections define approaches suggested for reaching the required improvements over the present level for each of these three major mission technical characteristics.

Specifically, in section~\ref{s:ant-syst} we present possible configurations for spaceborne antennas enabling aperture sizes comparable or even exceeding those of  ground-based EHT telescopes, $\sim$10--15~m in diameter and larger. In section~\ref{s:t-Rx} we discuss the current trends in developing receivers and other analogue electronics which approach quantum noise limits. Finally, sections~\ref{s:data-handl}, \ref{s:th-sync} and \ref{s:t-604} describe the issues of data handling at the required data rates, heterodyning and synchronisation, and baseline vector determination, respectively, which together define the achievable time-bandwidth product. As explained in section~\ref{s:sci-case}, we focus on the observing frequency range 
from 230~GHz to 1.2~THz. The lower end of this range is chosen as the basic operational band of the EHT, the higher end corresponds to the wavelength at which the required angular resolution would not necessitate baselines exceeding significantly the Lunar orbit. We emphasise that these parameters are chosen as a study case which would help to create a basis for future engineering in-depth studies. 
We anticipate that the next steps in developing the THEZA concept will address the mission analysis and design studies within the boundaries formulated herein.

\section{Antenna system}
\label{s:ant-syst}

For a space-borne radio telescope, its antenna is the main component defining the spacecraft dimensional and mass characteristics. The size of the telescope's aperture does matter a lot for its sensitivity. However, limitations of launch vehicles put very strict constraints on the size of payload. As of today, a practical limit of payload diameter for traditional placement under the nose fairing of launchers is about 4~m. This limit is likely to stay for several decades to come. Due to this restriction, space-borne radio telescopes are designed either as relatively small antennas (comparing to their Earth-based counterparts) or as deployable in orbit structures. The former approach is especially relevant for short-wave, millimeter and sub-millimeter telescopes since they pose stricter requirements for geometrical precision of antenna structures. Such the ``single piece'' design has been demonstrated successfully in many mm/sub-mm space-borne telescopes (see \cite{Gurvits+2021} for relevant references). It is also considered as a base design for several proposed space-borne radio interferometers and single dish systems (e.g., \cite{Roelofs+2019,Fish+2020,Wiedner+2021,Linz+2021}). There was one known case of attempting to design a 20-m class space-borne antenna capable for observations at 220~GHz to be launched as a full-size monolithic structure for the International VLBI Satellite project (\cite{Pilbratt-1991}, \cite[p.~399]{Hendrickx-Vis-2007} and B.Ye.~Chertok, 1990, private communication). This futuristic option assumed strapping a full-size 20-m radio telescope to the tank of the ``Energiya'' launcher as a replacement of the ``Buran'' spaceplane in its nominal launch configuration. However, this idea never proceeded beyond very preliminary considerations.

An alternative to launching a radio telescope antenna as a single-piece structure is its deployment in orbit. This approach was implemented in all three demonstrated to date Space VLBI systems, TDRSS, VSOP, and RadioAstron, see \cite{Gurvits-2020} for corresponding references. A deployable structure is adopted for the prospective Russian-led mm/sub-mm wavelength mission Millimetron \cite{Kardashev+2014}.
The diameter of such a deployable antenna is still limited by the space available under the launcher's nose fairing. 

\begin{figure}[h]
  \centering
\includegraphics[width=0.80\textwidth,angle=0]{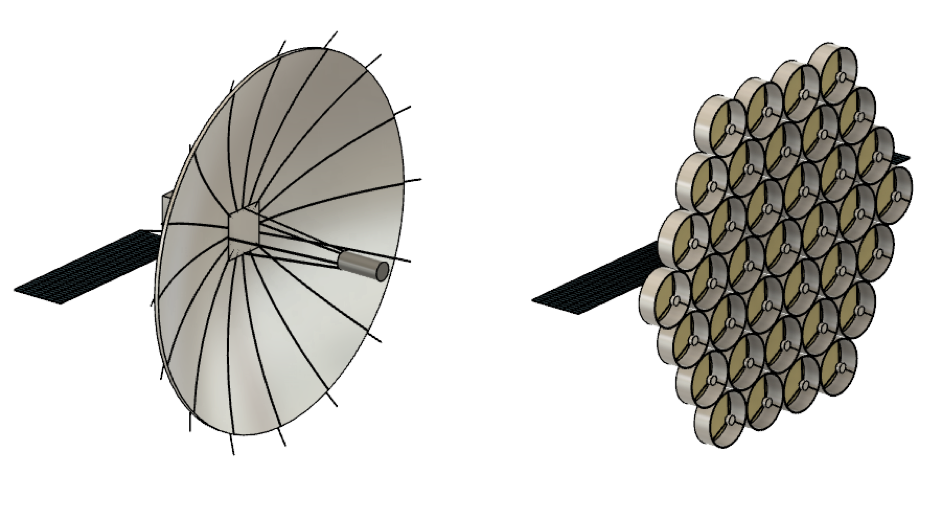}
\caption{THEZA spacecraft concept employing a single parabolic antenna (left) and a phased array (right).
}
   \label{f:ant-concept}   
\end{figure}

The issue of aperture size could be addressed in a very different way which was first suggested more than forty years ago in \cite{Buyakas+1979} -- an expandable space-borne radio telescope. This concept has two components: in-orbit assembly of a radio telescope and its infinite expandability. The former component was a subject of preliminary design study the Aerospatiale commissioned by ESA in the end of the 1990s. It considered a hypothetical next-generation Space VLBI radio telescope assembled on the International Space Station and then placed on an operational Earth orbit \citep{Gurvits-2000}. That study focused on assembling a traditional parabolic reflector of $\sim 30$~m diameter. An interesting component of that study was use of the European Robotic Arm (ERA)\footnote{https://www.esa.int/Science\_Exploration/Human\_and\_Robotic\_Exploration/International\_Space\_Station/European\_Robotic\_Arm, accessed 2021.07.23.} -- a robotic manipulator for operations outside the International Space Station (ISS). ERA was launched to the ISS in July 2021. 

In recent years, there has been a trend of developing ever larger space-borne optical and infra-red telescopes with large primary apertures, with the James Webb Space Telescope with a diameter of 6.5~m launched in December 2021, being the largest. To overcome size limitations, segmented self-similar primary mirrors assembled in Space are being explored by the California Institute of Technology and the University of Surrey in the Autonomous Assembly of a Reconfigurable Space Telescope (AAReST) project. Such as concept has previously shown the feasibility of building up to 100\,m sized mirror arrays and could hold huge potential for the future \cite{Hogstrom+2014IAC}. As it has been demonstrated, the use of self-similar octahedral cells as building blocks is efficient for creating large mechanical metamaterial structures \cite{cheung2012,Janett+Cheung-2017}. A NASA study \cite{Trinth+2017} proved this concept of automating the assembly of large-scale space structures and could offer a viable process for autonomous formation of space-borne radio telescopes.

\begin{figure}[t]
  \centering
\includegraphics[width=0.85\textwidth,angle=0]{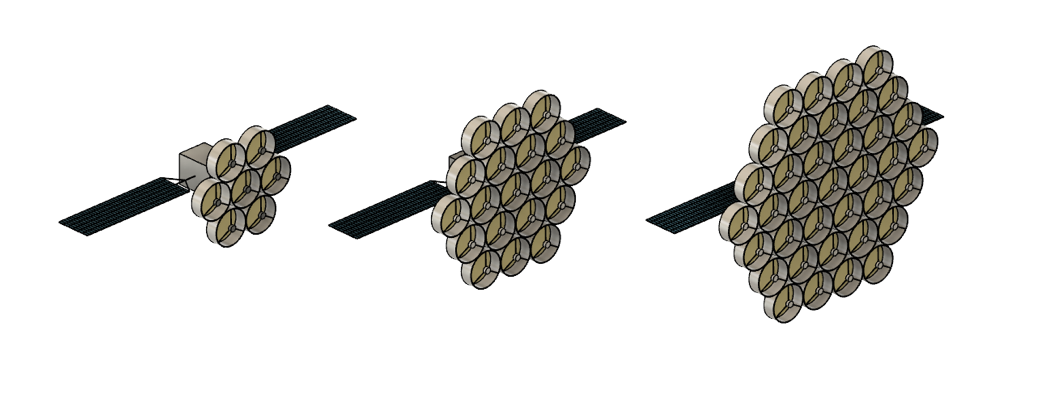}
\caption{An expandable configuration of THEZA aperture phased arrays with 7, 19 and 37 elements.
}
   \label{f:7-19-37}   
\end{figure}

Over the past two decades, aperture phased arrays made their headway into Earth-based radio astronomy at practically all wavelengths. The mm/sub-mm domain is no exception. As demonstrated by the Yuan-Tseh Lee Array (YTLA; also known as AMiBA, the Array for Microwave Background Anisotropy) \cite[and references therein]{AMiBA-2006}, built and operated on Mauna Loa (Hawaii, USA) by the collaboration between the Academia Sinica Institute of Astronomy and Astrophysics, the National Taiwan University and the Australia Telescope National Facility, an aperture array can operate successfully for millimetre radio astronomy studies. 

Implementing a phased array aboard the THEZA spacecraft would be a novel solution for space-borne radio telescopes, as this technique has not been utilised by past radio astronomy missions. The difference in Technology Readiness Level (TRL\footnote{https://www.nasa.gov/directorates/heo/scan/engineering/technology/technology\_readiness\_level, accessed 2021.11.17}) would impact the cost and time associated with developing both systems, which must be weighed against the advantages that the phased array configuration offers. As illustrated in Fig.~\ref{f:ant-concept} (right), a phased array would consist of a number of small, monolithic antennas arranged in a planar configuration. Each monolithic antenna would require an individual set of receivers and electronics. The signals from each of these antennas can be combined to synthesise the response of a single, large reflector. A phased array radio telescope has a number of advantages over a large single antenna of the same size. The former imposes less stringent requirements for the spacecraft attitude control and stabilisation as the primary beam of a monolithic antenna is inversely proportional to the reflector diameter.

\begin{figure}[t]
  \centering
\includegraphics[width=0.59\textwidth,angle=0]{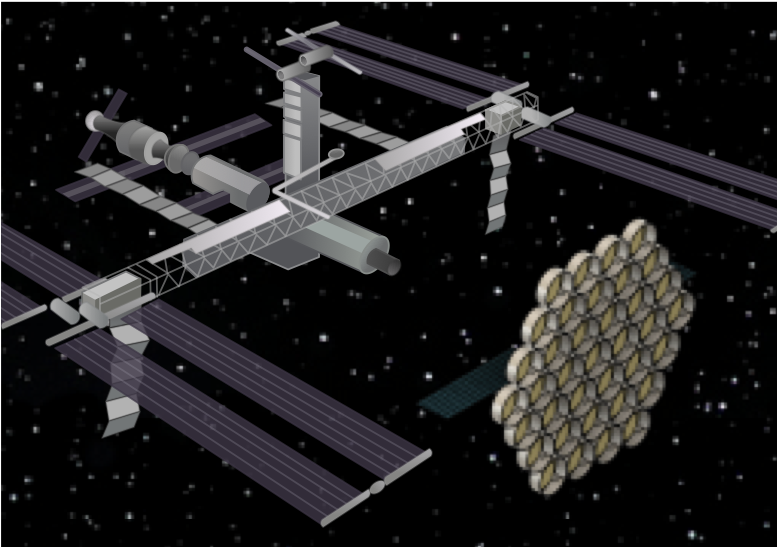}
\caption{A conceptual configuration of a THEZA spacecraft with an aperture phased array antenna assembled on an ISS-like orbital base.
}
   \label{f:issambly}   
\end{figure}

One might think of combining the principle design of an aperture array similar to YTLA with the in-orbit assembly approach. Each antenna element can be made of a size fitting for a launch configuration under a launcher nose fairing. Moreover, a package of individual reflectors can be arranged in a flat-pack launch configuration. A concept of flat-pack has been explored for unfolding solar arrays of the OrigamiSat design by NASA JPL and Brigham Young University \cite{Zirbel+2015SPIE}. A similar approach for a flat arrangement of a THEZA phased-array antenna elements might be considered for further in-depth study. In orbit, antenna elements are placed on a flat backup structure in the pattern similar to that of YTLA. This operation can be conducted by an ERA-like manipulator. The resulting aperture array telescope consisting of 7, 19, 37 or even lager number of antenna elements can reach effective diameter of 10--15~m or larger (Fig.~\ref{f:7-19-37}). The assembly of the space-borne aperture array might be considered as one of applications for a future multi-functional orbital base (e.g., Fig.~\ref{f:issambly}).

Since the element antenna in an interferometer is used mainly as a `photon collector', primary beam shape of it is not very critical. Higher aperture efficiencies can then be achieved by using non-parabolic shape in the reflectors. One example of these optimised shapes is Axially Displaced Ellipse (ADE), where theoretical aperture efficiencies exceeding 90\% can be achieved \cite{Prata+2003}. High-precision space-qualified solid mirrors with size reaching the limits of nose fairing of the launchers is already a proven technology. An example of this is the 3.5 m diameter primary mirror that was constructed of silicon carbide and was machined and polished to the required thickness (about 3 mm), shape, and surface accuracy by the Opteon company in Finland \cite{Sein+2003SPIE}.


\section{THEZA receivers and analogue electronics}
\label{s:t-Rx}


General qualitative requirements for radio astronomy and VLBI receivers can be described by three terms: sensitivity, stability and linearity. In addition, for VLBI it is also essential to use coherent receivers that preserve the phase of the signal of interest for subsequent correlation, once digitised. 
Space missions impose additional constraints on the receivers, such as the need to limit the mass and volume, as well as the cooling and consumed power. The current trend in radio astronomy developments is to push broadband Low-Noise Amplifiers (LNA) utilisation towards THz frequencies and to digitise the signal even up to the sky frequency whenever possible (homodyne receivers). Another trend is developments of receiving elements suitable for phased array feeds (PAF, \cite{Hampson+2012, Oosterloo+2010}) or focal plane arrays on one hand, and large-scale aperture phased arrays on the other hand, \cite{warnick2018phased}. In these systems, using the lowest noise technology is prohibitive due to the cost of fabrication and development of a large number of receivers. Therefore, they push towards commercial technologies where the cost-effectiveness, yield and repeatability of devices is much greater. The technology of choice are the Monolithic Microwave Integrated Circuits (MMIC) LNAs, that conveniently for our case are better suited for the higher frequencies. 

\begin{figure}[h]
  \centering
\includegraphics[width=0.49\textwidth,angle=0]{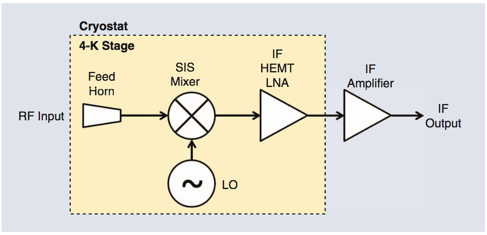}
\includegraphics[width=0.49\textwidth,angle=0]{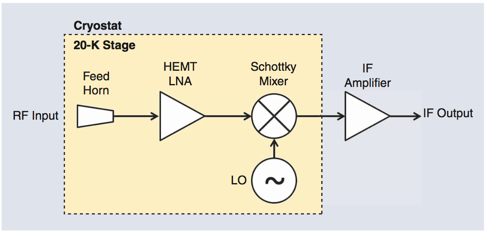}
\caption{A simplified block diagram of the heterodyne front-end architecture with ultra-cold mixers as first stage and low noise amplification at the intermediate frequency (IF) stage (left) and the heterodyne front-end architecture with direct amplification at radio frequencies (RF) and downconversion to IF (right) \cite{Cuadrado-Calle+2017a}.
}
   \label{f:frontend}   
\end{figure}

\begin{figure}[h]
  \centering
\includegraphics[width=0.59\textwidth,angle=0]{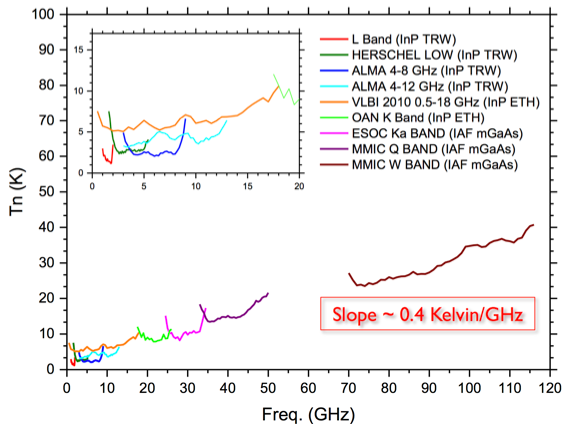}
\caption{Noise temperature (at $T_\mathrm{amb} = 15$~K) and frequency coverage of various amplifiers developed at Yebes Observatory. These amplifiers are used for either type of heterodyne receivers, some designs for both. The amplifiers above 20 GHz are used only in direct amplification heterodyne receivers, while the lower frequency ones (inset) are used also as IF amplifiers for the THz range.
}
   \label{f:Yebes-ampli}   
\end{figure}

The current status of the state-of-the-art developments in THz instrumentation allows us to estimate TRL of space-borne coherent THz detectors suitable for the THEZA concept. We differentiate between two types of THz receivers (Fig.~\ref{f:frontend}). The first type is the space-proven architecture of heterodyne receivers that uses super-conducting mixers as a first stage. Examples are the HIFI instrument on ESA’s Herschel mission \cite{deGraauw+2010} and the future HERO instrument on the Origins space telescope \cite{Wiedner+2021JATIS}, or the heterodyne instruments on the airborne observatory SOFIA \cite{Risacher+2018} and the future Millimetron Space Observatory \cite{Kardashev+2014}. The second type are also heterodyne receivers, but with direct amplification of the THz signal before downconversion to an intermediate frequency (IF). A recent development at the lower frequency end of the THz range is a joint upgrade for the ALMA Bands 2 and 3 \cite{Cuadrado-Calle+2017, Yagoubov+2020}. Figure \ref{f:Yebes-ampli} shows noise temperatures of different state-of-the-art low noise amplifiers developed at Yebes Observatory, that apply to both types of heterodyne receivers, e.g.~\cite{Tercero+2021} and references therein. Both architectures have pros and cons. The first type has high TRL (up to 9 depending on the components) as it has been already used for space missions \cite{Wiedner+2018} and will be the focus of the below discussion. The second type of receivers has very low TRL and will require further developments in order to cover rising sky frequencies into the THz range but could have clear advantages for space-borne instrumentation in terms of demand for spacecraft resources. A recent alternative is a graphene-based detector doped to the Dirac point that enables highly sensitive and wide-band coherent detection of signals from 90 to 700~GHz and, prospectively, across the entire terahertz range \cite{Lara-Avila+2019}, with a bandwidth up to 20~GHz. The new graphene detector would require less than a nanowatt of local oscillator power, enabling detectors with many pixels.

\subsection{Heterodyne receivers with cryogenic mixers}
\label{ss:hetero-Rx}

\begin{figure}[h]
  \centering
\includegraphics[width=0.59\textwidth,angle=0]{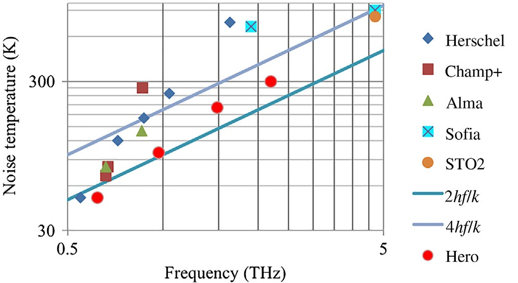}
\caption{Receiver noise temperature of different ground-based and space telescopes compared to the typical goal of $2 hf/k$ for SIS mixers and $4 hf/k$ for HEB mixers, where $h$ is the Planck constant, $f$ the frequency, and $k$ the Boltzmann constant \cite{Wiedner+2021JATIS}.
}
   \label{f:Rx-noise}   
\end{figure}

To access the THz range, this type of heterodyne receivers first downconvert the observed sky frequency to the GHz range, and the signal is immediately amplified at this IF stage. These devices require state-of-the-art components where the mixer, the local oscillator and the first amplifier are the most critical elements for low-noise performance systems. Figure \ref{f:Rx-noise} shows the noise temperature for different heterodyne receivers with ultra-cold mixers used in past, present and future telescopes \cite{Wiedner+2021JATIS}.

When these receivers are used in space-borne missions, it is necessary to take into account that they are very demanding on spacecraft resources and their consumption needs to be carefully controlled. The mixers need to be refrigerated at 4~K or even lower temperatures for optimal performance, using for example superfluid helium cryostats. Typically, only $\textless$ 100~mW are available for cooling power at 4~K for the spacecraft scientific payloads. The amount of cryogenic agent also limits the mission lifetime, typically planned for a five years lifetime. An alternative approach would be to use cryogen-free mechanical coolers for space missions to prevent the mission operation lifetime being limited by the amount of refrigerant \cite{Jones+1995}, only limited by wear degradation in this case. The non-selected mission SPICA \cite{Shinozaki+2020} planned to use a 4~K / 1~K-class Joule--Thomson coolers to cool the telescope and the thermal interface for the focal plane instruments, and two sets of double stage Stirling coolers to cool the telescope shield. The Millimetron Space Observatory will not have cryogenic liquids on board, it will instead be cooled passively by heat shields and actively by mechanical coolers \cite{Smirnov+2018}. Combined solutions will most probably be used in future space missions to alleviate the refrigerant requirement. An overview of the different cryogenic techniques and their operating temperature is given in \cite{rando-2010}.  


A review of the current state-of-the-art components for space-borne heterodyne receivers with ultra-cold mixers is presented in \cite{Wiedner+2018}. Below we briefly summarise the status of these components. A justifiable extrapolation of the current state of relevant technologies confirms the feasibility of the THEZA receiver capabilities.

\subsubsection{Mixers}
\label{sss:mxie}

At frequencies above 70~GHz, radio astronomy receivers have traditionally used Superconductor-Insulator-Superconductor (SIS) mixers configured as heterodyne detectors due to their ability to provide low system temperatures. They need to be refrigerated at 4~K for near-quantum noise performance. They work well at frequencies up to $700-800$~GHz with IF bandwidths reaching 12~GHz or higher with slight degradation of the system temperature. Examples are the Band 9 ALMA receivers \cite{Baryshev+2015}. Another recent receiver developed by NAOJ provides a very large bandwidth covering the $4-21$~GHz IF band \cite{Kojima+2020}.

For higher frequencies, Hot Electron Bolometers (HEB) mixers are used instead. These mixers need to be refrigerated at even lower temperatures, down to $\approx$ 0.3~K. Latest developments still provide modest IF bandwidths, up to 7.5~GHz on NbN-based mixers \cite{Krause+2018}. Promising results using MgB$_{2}$-based mixers show good receiver noise temperatures even at 20~K operation \cite{Novo-Chered-2017}.  

While SIS mixers require a local oscillator with 10~nW power for the pump, HEB mixers need even less, about 1~nW.

\subsubsection{Local oscillator sources}
\label{sss:lo}

The challenge for the coherent sources is that they need to be tuneable over a very wide frequency range, reach high frequencies and have low power consumption. In case of multi-pixel receivers (as for PAF receivers), they also need to provide enough power to pump multiple mixers.   

Recent developments are based on Schottky diode-based frequency multiplier  \cite{Mehdi+2017}, pumped by high-power GaAs amplifiers at the 3-mm W-band  \cite{Siles+2015}. An alternative research uses quantum cascade laser to be able to reach the highest THz frequencies, up to 4.7~THz \cite{Huebers+2018}], to detect the neutral oxygen OI line.

\subsubsection{Intermediate frequency cryogenic Low Noise Amplifiers} 
\label{sss:cryo}

Cryogenic amplifiers for space missions need to have high TRL and very low power dissipation. Requirements have been evolving from 10~mW for a simple receiver, to about 0.5~mW per IF chain for the multi-pixel Origins mission \cite{Wiedner+2021JATIS}. 

The lowest power dissipation till date has been obtained with cryogenic SiGe heterojunction bipolar transistors with good noise performance (5~K) but with limited IF bandwidth (e.g. 1.8~GHz \cite{Montazeri+2016a} and 4~GHz \cite{Montazeri+2016b}) due to the increase of noise at higher frequencies. On the contrary, they adapt easily to the lower frequencies. Alternatively, the InP HEMT (High Electron Mobility Transistor) technology is being used with larger bandwidths and the best noise results. This technology has been already space qualified for HIFI/Herschel \cite{Lopez-Fernandez+2003}. Further development was needed to demonstrate good performance at reduced power as shown by \cite{Wadefalk+2003} or more recently by \cite{Cha+2020}, achieving an ultra-low power dissipation of 112~$\mu$W. Other amplifiers are based on metamorphic GaAs HEMTs \cite{Tercero+2021, Abelan+2012}. These developments take advantage of the industrially established procedures for the GaAs but emulating the layer structure in the InP HEMTs. Some of these amplifiers\footnote{Low Noise Factory: https://www.lownoisefactory.com, accessed 2022.01.09} and transistors\footnote{Diramics: http://diramics.com, accessed 2022.01.09} are now commercially available.

Lastly, it is worth to mention the parametric cryogenic amplifiers that dissipate much lower power \cite{Vissers+2016} and have very low noise, close to the quantum limit \cite{Kutlu+2021}, but they easily saturate and require microwave pumping. Apart from axion search experiments \cite{Braine+2020, Crescini+2020}, these amplifiers still need to be demonstrated in ground-based receivers, therefore, currently only low TRL levels can be attributed to these devices.


\section{THEZA data transport and processing}
\label{s:data-handl}

A key component in THEZA data processing is the VLBI correlator. This equipment computes the coherence between the observed wave front as a function of relative delay and its time derivative(s) between all pairs of independently observing satellites and/or ground stations (``elements''). Proven implementations of this algorithm exist in hard-, firm- and software. However, none of the underlying technologies can scale to the necessary data processing rates required by THEZA. Apart from pure computational requirements, Space--Space or Space--Earth VLBI correlation opens up several areas of research in power consumption/computation, data transmission/storage, advanced fringe search- and tracking algorithms – taking into account the time-dependent geometry of Space--Space and Space--Earth baselines, as well as instrumental element-related and baseline-related amplitude and phase noise.

This section discusses in more detail the requirements and trade-offs to make for the THEZA correlator. 

\subsection{Organisation of raw VLBI data flow and storage}
\label{ss:data-logi}

VLBI correlation can be summarised as measuring and storing the amplitude and phase of electromagnetic waves at different points in space persistently, and create a Michelson--Morley style interference pattern -- let the signal interfere with ``itself'' -- afterwards. The quotes around ``itself'' are because of the fact that in VLBI the single signal is not split by a beam-splitter, rather it is the same wave front sampled at different positions in space.

In order to compute the interference pattern afterwards, it is necessary, as in the Michelson--Morley experiment, to bring the wave front data sampled by different elements together in a central location. At this point, a computer having access to the data streams of the elements can compute the correlation function, or coherence.

This VLBI correlation in itself is a moderately simple algorithm. The problem with VLBI correlation lies in scaling: the dimensions of the VLBI correlator system as a whole are governed by scaling laws as listed below:

\begin{enumerate}
\item \textit{the expected average number of observing elements:} the number of interferometer pairs to compute scales as number of elements squared;
\item \textit{the data sampling rate:} relating to the observed bandwidth per element, the correlator computational load scales linearly with this number;
\item \textit{the sample bit width:} combined with the data sampling rate it yields the observing data rate per element; the correlator computational load does not typically scale with this number, the data transmission and storage volumes on the other hand do, linearly;
\item \textit{the maximum supported interferometric ``fringe'' search window:} the correlator computational load scales as $\log_2(n)$ with $n$ being the number of points in the search window.

\end{enumerate}

Current cm- and mm-wavelength VLBI observations observe fractional bandwidths ranging from 16\% (256 MHz at 1.6 GHz) to 1.6\% (4 GHz at 230 GHz). For a VLBI imaging instrument to be useful, the fractional bandwidth observed should be in this range. Extrapolating this to THEZA, observing at THz frequencies, a fractional bandwidth of just 1.6\% already translates to 16 GHz. For critically sampled VLBI data –  two bits per sample Nyquist sampled, dual polarization – this represents a data rate of 128 Gigabit per second (16 GiB) per element.

To put the instantaneous computational requirement for THEZA in perspective: the current (2021) $\sim 1100$-core CPU cluster at the Joint Institute for VLBI (JIVE) handles 512 MHz of bandwidth for ten elements in real time. The linear scaling relation of point 2. above results in a ten-element THEZA correlator requiring at least 32 times that, i.e. $\sim 36\,000$ cores. Whilst not prohibitively large, it would put it in a contemporary TOP500 list of supercomputers. Experiments with GPU- and FPGA- based accelerators have indicated that the obtained gain in performance does not balance well against the invested effort, as well as having difficulty in being capable of supporting the flexibility required to address the complexity of the THEZA problem (i.e. correlations involving at least one Space-based element), especially for low number of station correlations.

A key word in the previous paragraph is \textit{real-time}. Computational requirements can be lowered by allowing slower-than-real-time correlation, e.g. by using less CPU cores and/or running at lower clock speeds to conserve energy. The trade-off to be made here is that the elements' data needs to be recorded on a persistent medium so as to allow for offline correlation and that the duty cycle of observations is small enough to finish correlation before the new observation cycle begins, such that the available storage can be reused.

The requirements of the storage, transmission, and compute subsystems of a VLBI correlation system are very much linked to each other but can be flexibly exchanged within the constraint boundaries. If the storage system near the receiver is capable of keeping up with the raw data speed and can hold the raw data volume, a slower-than-real-time transmission system from the element to the central location can be matched to the available compute performance. Alternatively, the computing system can be matched to the available transmission speed. If the transmission and computing subsystem can sustain real-time performance with a delay and delay rate fringe-search window large enough to accommodate the a priori accuracy of the baseline vector determination subsystem, the storage subsystem could be removed. To illustrate: the relation between required fringe-search window in number of frequency points to correlate $N_{\rm{fft}}$ for an a priori positional accuracy $\Delta x$ in meters is $N_{\rm{fft}} = x \cdot \Delta\nu / c$, where $\Delta\nu$ is the channel bandwidth (in Hz) and $c$ the speed of light in vacuum. The formula assumes standard VLBI observing parameters: a Nyquist-sampled band at two bits per sample. Just as in many other major specification parameters of the THEZA concept, a trade-off between the baseline vector determination precision (section~\ref{s:t-604}) and parameters of the data processing system should be addressed at the next step of the study, the design phase of the mission based on the THEZA concept.

As a rule of thumb, the relative speeds of the three subsystems should not differ by more than a factor of two to allow a sufficiently high duty cycle on the instrument as a whole, although if the storage volume is large enough: 
$$ N \cdot 4 \cdot \Delta\nu \cdot \tau_\mathrm{obs} \cdot \epsilon \le V_\mathrm{storage}$$ subject to the condition $$\epsilon \le \frac{\tau_\mathrm{corr}}{\tau_\mathrm{obs}}$$ with $4 \cdot \Delta\nu$ being the \textit{observing data rate} (Nyquist sampled bandwidth $\Delta\nu$ at $2$ bits/sample), $\tau_\mathrm{obs}$ the \textit{observation duration}, $\epsilon$ the \textit{duty cycle} of the instrument, $\tau_\mathrm{corr}$ the \textit{time taken to correlate the recorded data}, and $N$ the \textit{number of elements in the array}, in principle any $0 < \epsilon \leq 1$ might be supported.




\subsection{Data transport}
\label{ss:data-trans}

For any VLBI system, data transport is a major issue defining the system architecture and specifications. This is especially relevant for a Space VLBI systems. In both implemented to date Space VLBI missions, VSOP \cite{Hirabayashi+1998} and RadioAstron \cite{Kardashev+2013}, raw VLBI data were downlinked from orbit to a dedicated network of Earth-based data acquisition stations at Ku-band ($\sim$14~GHz). For the VSOP, the network consisted of five Earth-based stations equipped with $\sim$10-m-class antennas. RadioAstron operated with two data acquisition stations with antennas of 22~m and 43~m. In both missions, the data were transmitted through the Ku-band down-link channels at the rate of 128~Mbit/s. This was a practical limit at the time of the mission design in the 1990s. Importantly, in both cases, the VLBI data streams were downlinked in real time, without storing them aboard the spacecraft. At this point of the THEZA concept development, it is hard to definitively decide whether the mission should operate in the mode of real-time data transport to the processing facility. However, given the expected amount of raw data involved in VLBI imaging experiments, the intermediate storage (e.g., on board a THEZA spacecraft) appears to be unrealistic.  

In order to achieve its science objectives, THEZA must provide a VLBI data streaming from each of its interferometric stations comparable to that used in the Earth-based EHT system. In practical terms, it means that the data rate per THEZA station must be of the order of 100~Gbit/s. This value exceeds the currently demonstrated in space-borne systems data rate, both in radio and optical systems. The latter apparently provide the highest achieved to date rate at the level of several Gbit/s over distances of the order of $10^{4}$~km \cite{Bohmer+2012,Heine+2015}. It is reasonable to expect that the ongoing studies of advanced optical communication systems would demonstrate the data rate of the order of terabit per second \cite{Hauschildt+2019SPIE} in the coming decade. Depending on the overall THEZA configuration, the data transport system should provide only inter-satellite link (the case of Space-only baselines) with in-orbit correlation, or enable data down-link if the system involves both space-borne and Earth-based telescopes. The latter option would allow easing requirements for the correlation data processing at the expense of increased load on the data down-link system. The trade-off between these two data transport and correlation options is likely to be one of the main contributors in the choice of the overall mission architecture.

\subsection{Data correlation architecture}
\label{ss:data-corr}

Section~\ref{ss:data-logi} discusses the VLBI correlation system, similar to how current ground-based VLBI arrays are observing and their data logistics are organized. For THEZA, which introduces one, or preferably several, space-borne elements, other architectures at different levels can be considered. A crucial feature of a VLBI correlator not mentioned before is that it produces (a \textit{lot}) less output data volume than it takes in.

Space-borne elements require a downlink. The design and implementation of a downlink scales worse than linear 
with increasing requirements on the link. Depending on the choice of correlation architecture, the requirements on the downlink may differ by more than several orders of magnitude between the options.

With more than one space-borne elements, the choice can be made to perform in-orbit correlation of the space-borne elements only. This reduces the downlink performance from $N$ $\times$ 16~GB per second to $N^2 \times n \times 8$ bytes per second (assuming the correlator produces 8-byte complex numbers). To illustrate the magnitude of the impact: for an $N=5$ element space-borne array and a large fringe search window, $n = 4096$, this leads to a downlink requirement drop by a factor of 10$^5$.

In such a case, the transmission between telescope and storage systems on board of the space-borne elements becomes subject to the $\epsilon$ and $V_\mathrm{data}$ constraints mentioned in section \ref{ss:data-logi} as the space-borne array now becomes its own VLBI array. In this situation an interesting extra area of research will be the investigation of distributed correlation architecture(s), seeking to remove the central location bottleneck. A centralised correlation subsystem needs to be able to support the aggregate data rate and/or volume of all the elements in the array. A distributed architecture would naturally eliminate this Single Point of Failure (SPOF), at the cost of higher intra-element transmission capacity \cite{rajan2013ac2}. If the elements' data is not sent to a central location, some duplication factor $>$ 1 is required to be able to form the $N^2$ interferometer pairs. As long as this duplication factor $< N$, the requirements on the intra-element links can be relaxed.

A space-borne VLBI array operating at THz frequencies will provide scientific results impossible to obtain from the ground. However, the scientific performance of a VLBI array, and more specifically the imaging capability at different physical scales and the imaging efficiency (how long the array needs to observe to arrive at a threshold performance) depends not just on the number of interferometer pairs but more so on the distribution of physical baseline lengths sampled during an observation.

The addition of space-borne elements in an observing frequency range available to existing ground-based VLBI arrays such as the EHT will greatly enhance the imaging performance compared to when both sets of elements operate as individual ground-based and space-borne arrays.

If sufficient downlink capacity can be organised to form a combined ground--space VLBI array this will not only increase the imaging performance of both arrays but also lower computational energy efficiency constraints compared to the stringent space-borne energy efficiency requirements, but also lower the storage and intra-element transmission requirements -- ground-based storage is easier and intra-element transmissions are unnecessary in this scenario. Additionally, having access to the raw space-element data on the ground means that re-correlation, debugging, and essentially any other reprocessing of the raw data to look for unknown science is possible. An example of this is the reprocessing of ground-based data many years after it was observed to look for transient signals with specific time-frequency signatures that were discovered later (Fast Radio Bursts, FRBs).



\subsection{Post-correlation processing}
\label{s:post-proc}

Given the extensive experience with VLBI post-correlation processing and ongoing efforts, there are no foreseeable bottlenecks for this step. Mature software is readily available, and active development ensures that specific requirements can be added. Dedicated efforts are underway to explore the high-performance computing domain, from which space-based VLBI will definitely benefit.

The main challenge arises from the acceleration of the interferometric space-only or space--ground baselines. This causes the correlated phases on baselines to have residual errors that depend quadratically on time, while the step to correct the errors in traditional Earth-based VLBI correlators assumes a linear dependence. The PIMA software package \cite{Petrov+2011} can fit for a quadratic term in time. Correction for a quadratic frequency dependency has been implemented in the CASA package \cite{McMullin+2007ASP}, this can easily be extended to the time domain.

Additional aspects to be aware of are meta-data collection and formatting, and the fact that the calibrator sources will be resolved at these baselines and frequencies. Older software assumes a point source calibrator, newer packages can include a source model to overcome this caveat.


\section{THEZA heterodyning and synchronisation}
\label{s:th-sync}

There are two possible approaches for achieving coherence for a space-based radio interferometer in the THz range of observing frequencies: one where each spaceborne  telescope has its own local frequency standard, or another one with the telescopes linked by a distributed frequency reference.

Hydrogen masers have been used as frequency references in several space missions, e.g. the Radio Astron mission, and are commonly used as reference clocks in VLBI. However, at the upper end of the THEZA frequency range (1.2\,THz), the system would suffer from degraded sensitivity due to coherence losses when using H-masers. If we optimistically model a modern active H-maser as having an ADEV of $\sigma_y(\tau) \approx 1e-13/\tau$, the expected correlation loss at 1.2 THz at 1 second will already be more than 50\% for integration times beyond 100\,s (see \cite{TMS-2017}, p.~436). 

Cryogenic Sapphire Oscillators (CSO) offer better short term performance than hydrogen masers and have been studied for use in VLBI at sub-mm wavelengths \cite{Rioja+2012AJ}. A recent CSO achieves $\sigma_y(\tau) < 10^{-15}$ for $1\,{\rm s} \leq \tau \leq 2,000\,{\rm s}$ \cite{Giordano+2016JPh}. However, space qualified CSO's are not available yet.

The use of a distributed reference frequency would shift the Allan deviation requirement from being on the clocks, to being on the distribution of the clock signal between the space elements. The stability of the distributed clock then only needs to be sufficient for coherence on timescales of the round-trip time between the in-orbit space observatories, a fraction of a second. Free space optical frequency links would also deliver very accurate relative Doppler measurements on each link. Optical links would also be an interesting candidate for the required high speed data transfer between the interferometer elements and the correlator facility (in orbit or on the ground). Such the technology has been demonstrated by the NASA’s Lunar Atmosphere and Dust Environment Explorer (LADEE) mission in collaboration with an ESA ground tracking station, \cite{Arnold+2014} and references therein.

Optical distribution of a reference frequency can easily outperform the stability of the microwave based references discussed above. Free space optical links to a satellite could offer a stability of better than $10^{-16}$ in one second \cite{2021arXiv210312909G}, which promises no coherence loss at 1.2\,THz for integration times of up to 10,000\,s, and 3\% at 100,000\,s. The optical reference distribution could either be between a ground station and the satellite segment, or amongst the stations in orbit themselves.

\section{THEZA baseline vector determination}
\label{s:t-604}


The precise knowledge of the absolute and relative locations of the phase centers of the THEZA telescopes, their velocities and accelerations is a requirement for the interferometric data processing. Ideally, the baseline vector connecting the respective phase centers should be determined with a precision of the order of the wavelength, which comes down to sub-mm for terahertz frequencies. This requirement is unrealistic with the current (and near-future) tracking capabilities and precise orbit determination methods. However, as well known from the practice of interferometry, and VLBI in particular, processing of interferometric measurements allows to mitigate this problem by searching for the interferometric response within wide enough windows of delay, delay rate and sometimes delay acceleration, enabling relaxation of requirements for the precision of {\sl a priori} knowledge of baseline vector, its velocity and acceleration, respectively. The determination errors of larger baselines put a higher demand on such processing and vice versa. Thus, a trade-off has to be made between instrumentation and methodology dedicated to precise orbit and baseline vector determination and the subsequent time derivatives, as well as capabilities/possibilities for advanced processing of interferometric observations.  

For precise orbit determination of Low-Earth Orbiting (LEO, $200-1500$~km altitude) and Medium-Earth Orbiting (MEO, around $20\,000$~km altitude) satellites, accuracies at the cm level are achieved when making use of high-quality, dual-frequency Global Navigation Satellite Systems (GNSS) receivers \citep{bock2011a,peter2017a,ijssel2015a,johnston2017}. For baselines around 200~km, a precision level as good as 0.5~mm has been achieved post-facto, making use of sophisticated dynamic force modelling and parameter estimation schemes \citep{kroes2005a,mao2019b,barneveld2012}. The force modelling typically includes high-precision knowledge of Earth's static and time-variable gravity field and third-body perturbation (luni-solar and planetary ephemeris). In addition, non-gravitational forces need to be modelled or measured precisely as well. These forces are due to solar radiation pressure and -- when flying low -- atmospheric drag. They can be derived either directly from observations collected by on-board accelerometers or by precise modelling. The latter requires precise knowledge about satellite geometry, surface properties (e.g. reflection) and satellite attitude (e.g. as can be derived from observations taken by star cameras). Furthermore, precise knowledge is required of the exact location of the center-of-mass of the satellite and the location of the phase center of the GNSS antenna needs to be precisely characterised.

It is fair to assume that current capabilities of precise orbit and baseline determination are applicable to satellites that form a THEZA constellation as well. However, several aspects need to be carefully considered and studied. For satellites flying at altitudes above the GNSS satellites, the tracking geometry is quite different and less favourable than for LEO satellites. Moreover, parameter estimation schemes for the best possible baseline determination typically include so-called integer phase cycle ambiguity fixing which has worked very well for relatively short baselines up to a few hundreds of kilometers for the controlled GRACE tandem \citep{kroes2005a,barneveld2012} but is challenging for longer and also for more dynamic baselines such as between the lower and higher flying Swarm satellites \citep{mao2019b}. 


\begin{figure}[t]
  \centering
\includegraphics[width=0.80\textwidth,angle=0]{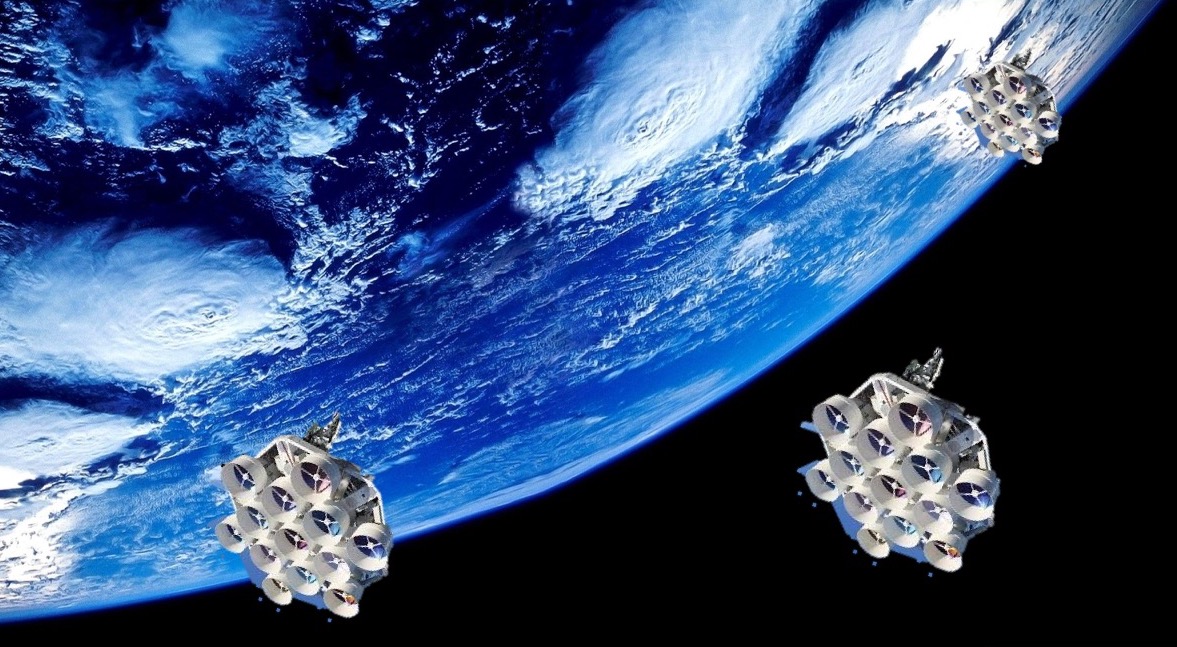}
\caption{Artist's impression of a three-element version of the THEZA concept.
}
   \label{f:THEZA-art}   
\end{figure}

The development of novel technologies and algorithms for very precise inter-satellite ranging is very promising. For GRACE Follow-On, a Laser Ranging Instrument (LRI) was implemented as a test together with K-band radio-wave ranging (KBR) system. The LRI provides inter-satellite ranges with a precision of about 1~nm/$\sqrt{Hz}$ at 100~mHz \citep{abich2019}. Also for GRACE Follow-On, the intersatellite distance is of the order of 200~km and is controlled. For LISA, relative displacements need to be measured at the level of pm for a baseline of the order of hundreds of thousands or even millions of kilometers \citep{danzmann2017}. Precise baseline determination for a THEZA satellite constellation will most certainly benefit from embarking high precision inter-satellite ranging instruments. Furthermore, the relative kinematics (position, velocities and acceleration) of the satellite array can also be estimated using co-operative communication within the satellite array and statistical algorithms, which could further improve the accuracy of the time-varying baseline vectors \cite{rajan2019}.

\section{Conclusions}
\label{s:concl}

The THEZA concept presented in this paper addresses a diverse set of scientific objectives which amends the science objectives of the ESA Voyage 2050 White Paper \cite{Gurvits+2021}.  The prime goal of the concept is creation of an interferometric facility able to make a major step in sharpening angular resolution in imaging observations by at least an order of magnitude over the best parameters achieved in the Earth-based EHT and Space VLBI RadioAstron observations. We consider one specific science case, studies of photon rings around supermassive black holes as the one defining the major target specifications of the concept: the angular resolution of about 1~$\mu$as, observing frequencies between 220~GHz and 1.2~THz, and interferometer antenna size 10$-$15~m in diameter. As argued in this paper, achieving the goal of the concept is only possible by placing interferometer elements in space -- no matter what option of increasing the angular resolution is chosen, increase of the VLBI baseline or shortening of the observing wavelength. Due to this reason, we believe that a space-borne interferometric system as presented in this paper is an inevitable step in the progress of observational astrophysics: earlier or later, a system similar to the THEZA conceptual three-element space-borne interferometer depicted in an artist's rendering form in Fig.~\ref{f:THEZA-art} will become a reality. We also addressed some, certainly not all, engineering challenges in creating an operational space-borne sub-millimeter interferometer. In particular, we presented the case for in-orbit assembly of a large telescope. We also provided brief summaries of the current status of other key components of the THEZA concept -- receivers, sub-systems for digital data handling and processing, heterodyning and synchronisation, and baseline vector estimates. All presented considerations should be considered as potential starting points for future in-depth engineering studies. 

\section{Acknowledgements}
\label{s:ackno}

The authors are grateful to the anonymous reviewers for valuable comments and suggestions. YYK and AVP were supported in the framework of the State project ``Science'' by the Ministry of Science and Higher Education of the Russian Federation under the contract 075-15-2020-778. JD is supported by a Joint Columbia/Flatiron Postdoctoral Fellowship. Research at the Flatiron Institute is supported by the Simons Foundation. JD is supported by NASA grant NNX17AL82G.





\bibliographystyle{elsarticle-num}
\bibliography{biblio-THEZA}







\end{document}